\documentclass[prl,aps,amsfonts,amsmath,amssymb,nofootinbib,twocolumn,superscriptaddress]{revtex4}

\usepackage{graphicx}
\usepackage{bm}
\usepackage{hyperref}
\usepackage{IEEEtrantools}
\usepackage{xcolor}

\usepackage{natbib}
\usepackage{float}

\restylefloat{table}

\begin{document}

\title{Probing charge pumping and relaxation of the chiral anomaly in a Dirac semimetal}

\author{Bing Cheng}
\affiliation{Department of Physics and Astronomy, The Johns Hopkins University, Baltimore, Maryland 21218, USA}

\author{Timo Schumann}
\affiliation{Materials Department, University of California, Santa Barbara, California 93106-5050, USA}

\author{Susanne Stemmer}
\affiliation{Materials Department, University of California, Santa Barbara, California 93106-5050, USA}

\author{N. P. Armitage}\email{npa@jhu.edu}
\affiliation{Department of Physics and Astronomy, The Johns Hopkins University, Baltimore, Maryland 21218, USA}

\date{\today}

\maketitle

\textbf{The linear band crossings of 3D Dirac and Weyl semimetals are characterized by a charge chirality, the parallel or anti-parallel locking of electron spin to its momentum.   Such materials are believed to exhibit a ${\bf E} \cdot {\bf B}$ chiral magnetic effect that is associated with the near conservation of chiral charge\cite{NPA18}. Here, we use magneto-terahertz spectroscopy to study epitaxial Cd$_3$As$_2$ films and extract their conductivities $\sigma(\omega)$ as a function of ${\bf E} \cdot {\bf B}$.  As field is applied, we observe a remarkably sharp Drude response that rises out of the broader background.  Its appearance is a definitive signature of a new transport channel and consistent with the chiral response, with its spectral weight a measure of the net chiral charge and width a measure of the scattering rate between chiral species\cite{burkov2018dynamical}. The field independence of the chiral relaxation establishes that it is set by the approximate conservation of the isospin that labels the crystalline point-group representations.
 }

The linear band crossings of 3D Dirac and Weyl semimetals are characterized by a charge chirality, the parallel or anti-parallel locking of electron spin to its momentum.   Such materials are believed to exhibit a ${\bf E} \cdot {\bf B}$ chiral magnetic effect that is associated with the near conservation of chiral charge\cite{NPA18}. Here, we use magneto-terahertz spectroscopy to study epitaxial Cd$_3$As$_2$ films and extract their conductivities $\sigma(\omega)$ as a function of ${\bf E} \cdot {\bf B}$.  As field is applied, we observe a remarkably sharp Drude response that rises out of the broader background.  Its appearance is a definitive signature of a new transport channel and consistent with the chiral response, with its spectral weight a measure of the net chiral charge and width a measure of the scattering rate between chiral species\cite{burkov2018dynamical}. The field independence of the chiral relaxation establishes that it is set by the approximate conservation of the isospin that labels the crystalline point-group representations.

Some of the most remarkable demonstrations of topological states of matter come through their response to electromagnetic fields.   Quantum Hall systems show quantized Hall resistances that are precise to better than one part in a billion \cite{KlitzingPRL1980} and topological insulators are characterized by a quantized magnetoelectric effect \cite{qi2008topological,essin2009magnetoelectric,wu2016quantized,armitage2019matter}.  Weyl (WSM) and Dirac semimetals (DSM) are states of matter in which conduction and valence bands touch and disperse linearly around pairs of nodes in momentum space \cite{abrikosov1970possible, nielsen1983adler, volovik2003universe,murakami2007phase,PhysRevB.83.205101,yang2011quantum,young2012dirac,NPA18}.  Each node is distinguished by its chirality e.g. whether the spin of a massless (linearly dispersing) particle is oriented parallel or anti-parallel to its momentum.  Roughly speaking Dirac systems can be considered as two copies of  Weyl systems, where at each node there are two sets of the linearly dispersing bands with opposite chiral charge.   The copies are distinguished by a point-group index or ‘isospin’ degree of freedom ($\Uparrow, \Downarrow $) that labels the crystalline point-group representations \cite{Cd3As2_LDA_2013,Parameswaran14a}.   These four-fold degenerate 3D linear band crossings in DSMs are protected by lattice point group symmetries and are stable as long as the symmetries are respected.  The quasiparticles near the touching points can be described by the relativistic Dirac Hamiltonian: {$\pmb{H}$ = $\eta$$v_F$$\pmb{\sigma}$$\cdot(\pmb{k}\pm$$\pmb{K_D}$)}, where $\eta$ = $\pm$ 1 represents the chirality degree of freedom and $\pm$ $\pmb{K_D}$ represent the valley degrees of freedom and location of the Dirac nodes along a high symmetry direction in momentum space \cite{Na3Bi_LDA_2012,Cd3As2_LDA_2013}.  

\begin{figure*}[t]
\includegraphics[clip,width=6.5in]{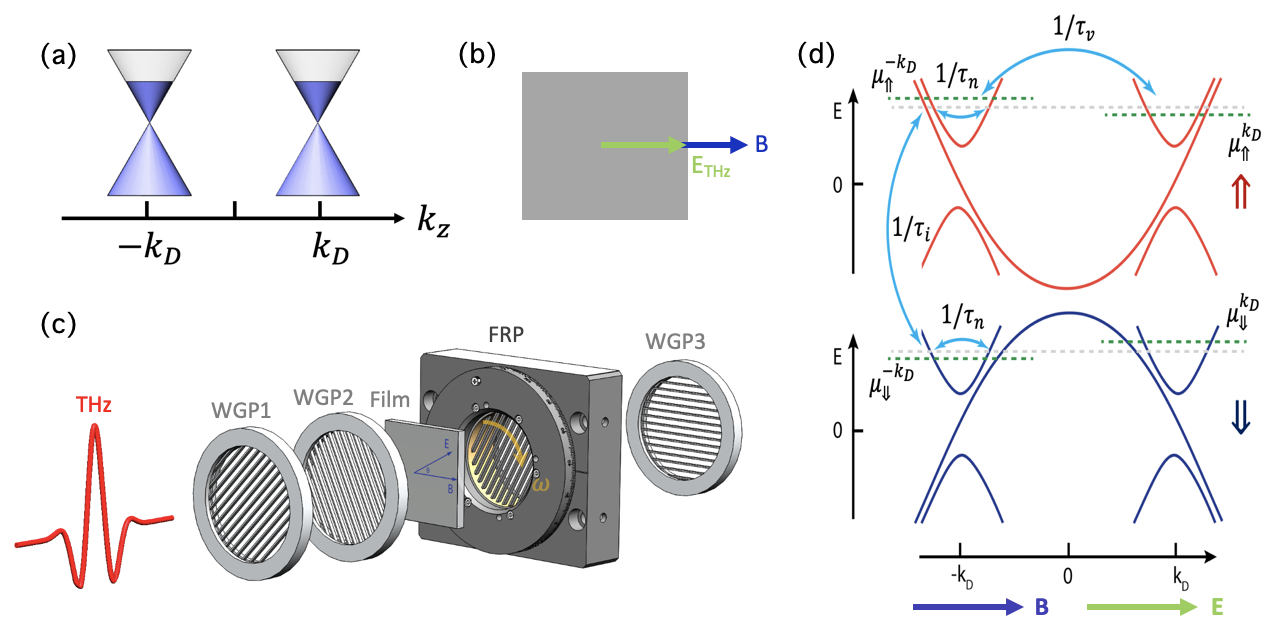}
\caption{{\bf Charge dynamics of the chiral anomaly in a Dirac semimetal and the experimental setup.}  {\bf (a)} Schematic illustration of the low-energy electronic structure of the Dirac semimetal Cd$_3$As$_2$. It hosts two 3D Dirac nodes located along the k$_z$ axis.{\bf (b)} The chiral anomaly is expected when the dc magnetic field and the THz electric field are co-aligned.  {\bf (c)} Schematic of the  time-domain magnetoterahertz spectrometer used to collect data. Wire grid polarizer 1 (WGP1) and WGP2 are used to produce linearly polarized terahertz pulse with ${\bf E}_{\mathrm{THz}}\parallel {\bf B}$ or ${\bf E}_{\mathrm{THz}}\perp {\bf B}$. A fast rotation polarizer (FRP) is used to modulate terahertz electric field by a frequency near 47 Hz. With WGP3 and lock-in amplifier, the complex transmission matrix can be determined through a single measurement to high precision \cite{morris2012polarization}. {\bf (d)} In a Dirac semimetal with ${\bf E}_{\mathrm{THz}}\parallel {\bf B}$, the 3D Dirac states will develop Landau levels which are dispersive along the direction of magnetic field. The zeroth Landau give the chiral current. A number of different relaxation rates control the charge dynamics.  $1/\tau_n$ is the intranode (normal) scattering rate,  $1/\tau_v$ is the intervalley scattering rate, and  $1/\tau_i$ in the internode scattering rate at the same momentum valley, but to the other isospin variety. }
\label{Fig1_new}
\end{figure*}

Despite being essentially metals, WSMs and DSMs can show distinct transport effects that are associated with the near conservation of chiral charge.   Unlike the quantum Hall effect, this ``chiral anomaly" exists in both the quantum and semiclassical transport limits \cite{nielsen1983adler,Son13a,Parameswaran14a,Burkov15a,burkov2018dynamical}. The phenomenon requires two important ingredients.  In the semiclassical limit, the first is a magnetic-field-induced coupling between the chiral and the total charge densities via Berry curvature.  This aspect is generic to systems with large Berry curvature.  The second is a nearly conserved chiral charge, which is a property unique to WSM and DSM systems.  The chiral charge is not exactly conserved in any real material, as the chiral symmetry is always violated by effects like nonlinear band dispersions, but such effects vanish in the low energy limit.  Therefore the near-conservation of the chiral charge is due to an emergent low-energy chiral symmetry.

Although the effect exists in both semiclassical and quantum transport regimes \cite{NPA18}, perhaps the most intuitive understanding of the effect can be achieved in the quantum limit.  Consider a magnetic field in the $z$ direction that bridges Dirac nodes separated in $k_z$.  As shown in Figs. \ref{Fig1_new}(d), due to the particular properties of massless Dirac fermions, a zeroth Landau level (LL) forms that connects one valley to the other either above or below E$_F$ depending on the relative direction of the magnetic field and the isospin distribution.  With the reasonable assumption (discussed more below) that inter-valley and isospin relaxation rates ($1/\tau_v$ and $1/\tau_i$) are slower than the intravalley rates ($1/\tau_n$) under the action of an ${\bf E} \cdot {\bf B}$ term, for a particular isospin, charge is pumped from one valley to the other.   For the opposite isospin, the sense of pumping between valleys is reversed, but note that due to the connectivity of the nodes of the 0th LL on isospin the velocity of the 0th LL at E$_F$ is always the same in the valley that charge is being pumped into.   This results in differences in valley ($\mu^{+k_D}_\Uparrow-\mu^{-k_D}_\Uparrow$) and isospin ($\mu^{+k_D}_\Uparrow-\mu^{+k_D}_\Downarrow$) chemical potentials and corresponding valley and isospin currents \cite{NPA18}.   These currents and population imbalances can be relaxed only by the relatively slow scattering between nodes of opposite chirality as compared to the normal intravalley scattering which dominates scattering across a node.   Inter-valley scattering ($1/\tau_v$) is slower than the normal scattering due to the large momentum transfer between nodes, and inter-isospin scattering ($1/\tau_i$) is suppressed by the same symmetry that protects the degeneracies.  We call the larger of $1/\tau_v$ and  $1/\tau_i$ the chiral relaxation rate $1/\tau_c$, which controls the build-up of valley charge imbalance.  Because the chiral charge is not precisely conserved e.g. it is pumped under the action of collinear electric and magnetic fields, the effect is referred to as the chiral anomaly.

\setlength{\parskip}{0em} 
A net valley or isospin population will lead to a chiral current, which relaxes at the lower rate  $1/\tau_c $ than a usual current and hence gives an enhanced dc conductance and negative longitudinal magnetoeresistance (NLMR).  NLMR has been observed in a number of DSM and WSM systems and was widely interpreted as a consequence of the chiral magnetic effect \cite{NPA18}.  However NLMR is not uniquely caused by this effect and it is clear that at least in the high mobility WSM system, the measurements were affected by the inhomogeneous current flow due to the large transverse magnetoresistances induced by magnetic field e.g. ``current jetting" \cite{NPA18,TaP_current_jetting_2017,pippard1989magnetoresistance,dos2016search,PHE_Na3Bi_PRX_2_2018}.  In lower mobility systems there are other possibilities including mobility fluctuations that may mix Hall and longitudinal responses in sufficiently thick films \cite{parish2003non,narayanan2015linear,Dirac_Timo_prb}.

\begin{figure*}[t]
\includegraphics[clip,width=6.7in]{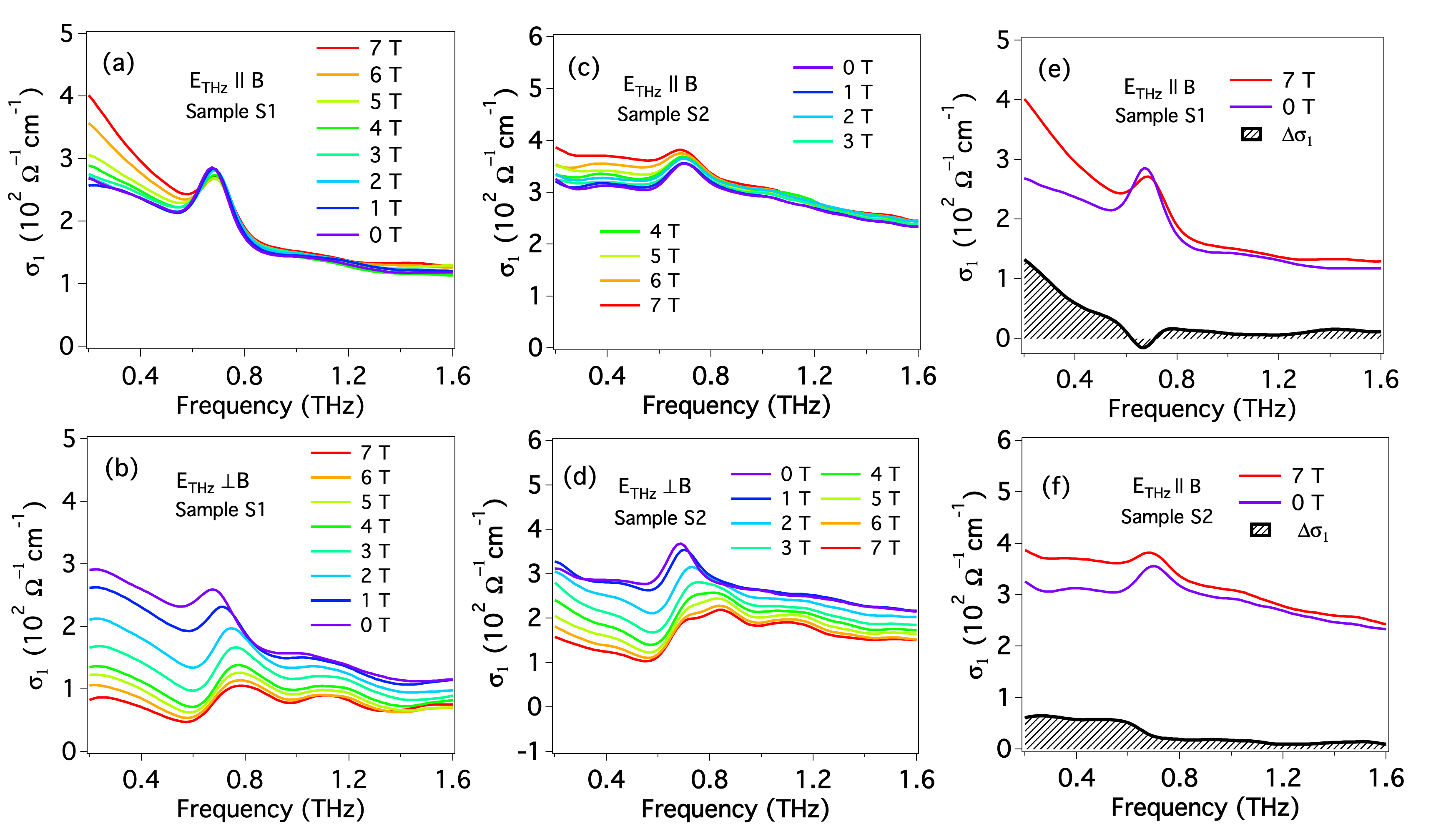}
\caption{{\bf Terahertz conductivity at different magnetic fields} {\bf (a)} ${\bf E}_{\mathrm{THz}}\parallel {\bf B}$ with $\bf{B} \parallel (\overline{1}10)$ for sample S1. Chiral anomaly leads terahertz conductivity $\sigma_1$ below 1 THz to be gradually enhanced by magnetic field. {\bf (b)} ${\bf E}_{\mathrm{THz}}\perp {\bf B}$ with $\bf{B} \parallel (\overline{1}10)$ for sample S1. The suppression of terahertz conductivity $\sigma_1$ is the signaure of postive magnetoresistivity which is generally observed in perpendicular magnetic and electric fields. {\bf (c)} ${\bf E}_{\mathrm{THz}}\parallel {\bf B}$ for  $\bf{B} \parallel (\overline{11}2)$ sample S2. {\bf (d)} ${\bf E}_{\mathrm{THz}}\perp {\bf B}$ for $\bf{B} \parallel (\overline{11}2)$ sample S2.  {\bf (e)} and  {\bf (f)} Comparisons of this 0 and 7 T data and their differences for samples S1 and S2. $\Delta\sigma_1$ is the intrinsic chiral conductivity from chiral anomaly. The highlighted grey area represents the strength of charge pumping effect and its width defines the chiral relaxation rate.}
\label{Fig20}
\end{figure*}

\begin{figure*}[t]
\includegraphics[clip,width=6.9in]{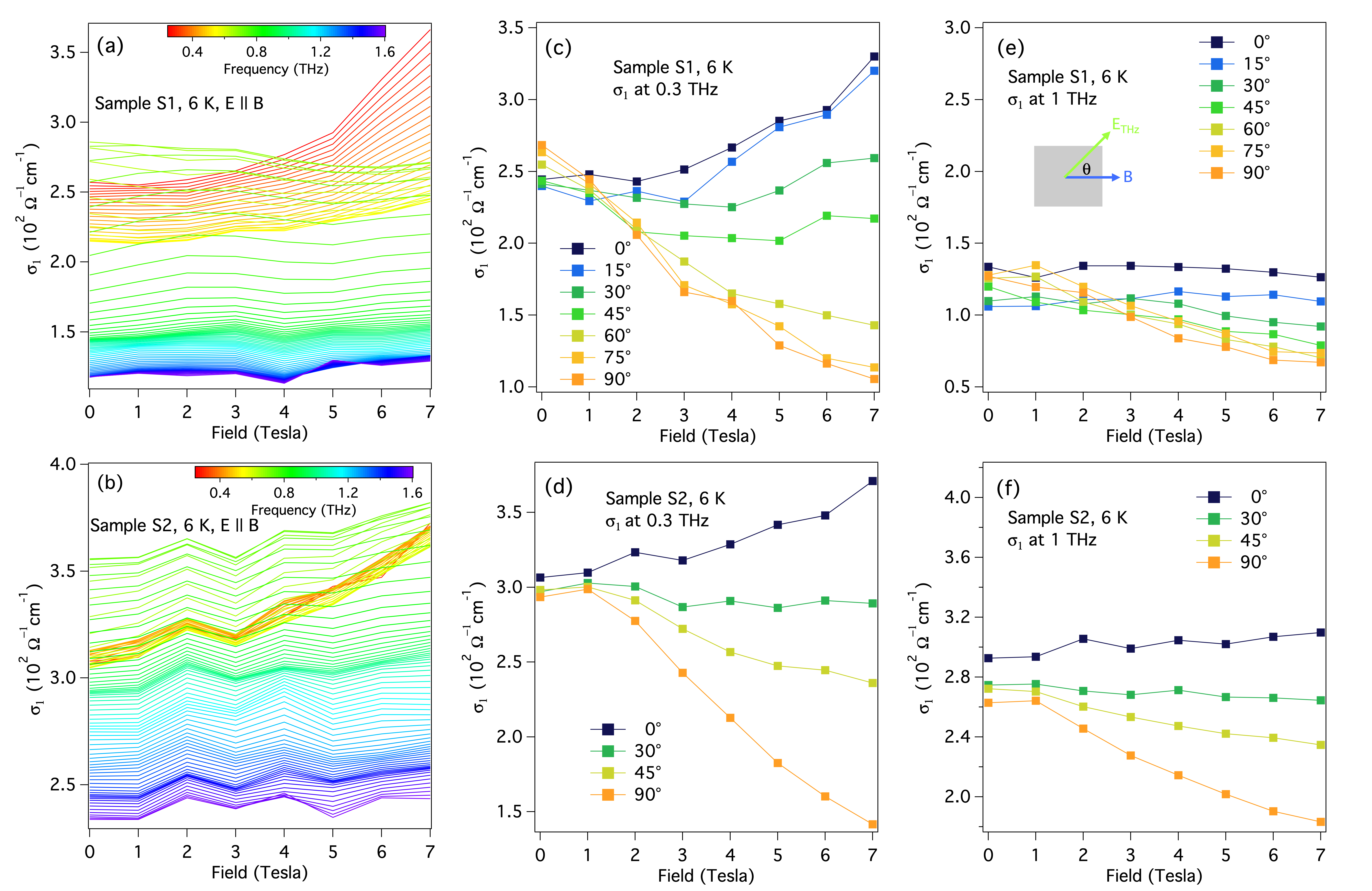}
\caption{ {\bf Terahertz conductivity as a function of magnetic field} Terahertz conductivity $\sigma_{1}$ at each frequency (see color bar scale) as a function of magnetic field of {\bf (a)} sample S1 and {\bf (b)} sample S2 with ${\bf E}_{\mathrm{THz}}\parallel {\bf B}$. Terahertz conductivity (at 0.3 THz) as a funcion of magnetic field under different terahertz polarization angles of {\bf (c)} sample S1 and {\bf (d)} sample S2. The configuration of polarization angle $\theta$ between terahertz electric field and magnetic field is shown by the schematic in Fig. 3(e). Terahertz conductivity (at 1 THz) as a funcion of magnetic field under different terahertz polarization angles of {\bf (e)} sample S1 and {\bf (f)} sample S2. All data were taken at 6 K.}
\label{Fig30}
\end{figure*}

Due to these controversies, experiments that can give conclusive information about the effect are desired \cite{levy2020optical,jadidi2019optical}.  As discussed above, a key parameter that governs the chiral anomaly is the chiral relaxation rate $1/\tau_c$. The most convincing way to characterize the intrinsic properties of the chiral anomaly would be to measure $1/\tau_c$ and $1/\tau_n$ directly. To date, most magnetotransport experiments of topological semimetals are performed with dc electric field for which it is hard to disentangle $1/\tau_c$ and $1/\tau_n$. In this work, we use magneto-terahertz spectroscopy as shown in Figs. \ref{Fig1_new}(c), to study the high-quality epitaxial thin films of Dirac semimetal Cd$_3$As$_2$.  Cd$_3$As$_2$ is an ideal material for this investigation as it simply presents two quadruply degenerate near-E$_F$ Dirac nodes that sit along the $k_z$ axis that are protected by a C$_4$ symmetry [Figs. \ref{Fig1_new}(a)].  High quality (112) oriented Cd$_3$As$_2$ films have been grown via molecular beam epitaxy \cite{Timo_CdAs_growth_16}.  Depending on growth parameters different E$_F$s can be achieved (E$_F$ = 46 meV and 118 meV for samples S1 and S2 respectively.  See the Supplementary Information (SI) for these estimates).  E$_F$ is generally lower in these films than in even good high single crystals \cite{akrap2016magneto,Cd3As2_phonon_2016}.  Performing frequency-dependent conductivity experiments offers the opportunity to extract $1/\tau_c$ and $1/\tau_n$ directly. We measured two Cd$_3$As$_2$ films and extracted their field-dependent terahertz conductivity. For the films with lower E$_F$, we found the emergence of a narrow Drude-like peak at low frequency when ${\bf E}_{\mathrm{THz}}\parallel {\bf B}$.  The appearance of this peak in a restricted low frequency range is the manifestation of a new transport channel.   Its systematic dependencies on frequency and field are in precise agreement with expectations of the chiral anomaly and the chiral Drude response in a Dirac semimetal.   An additional advantage of these contactless THz measurements is that they avoid any artifacts associated with inhomogeneous current paths that have plagued dc experiments.

\begin{figure*}[t]
\includegraphics[clip,width=7.2in]{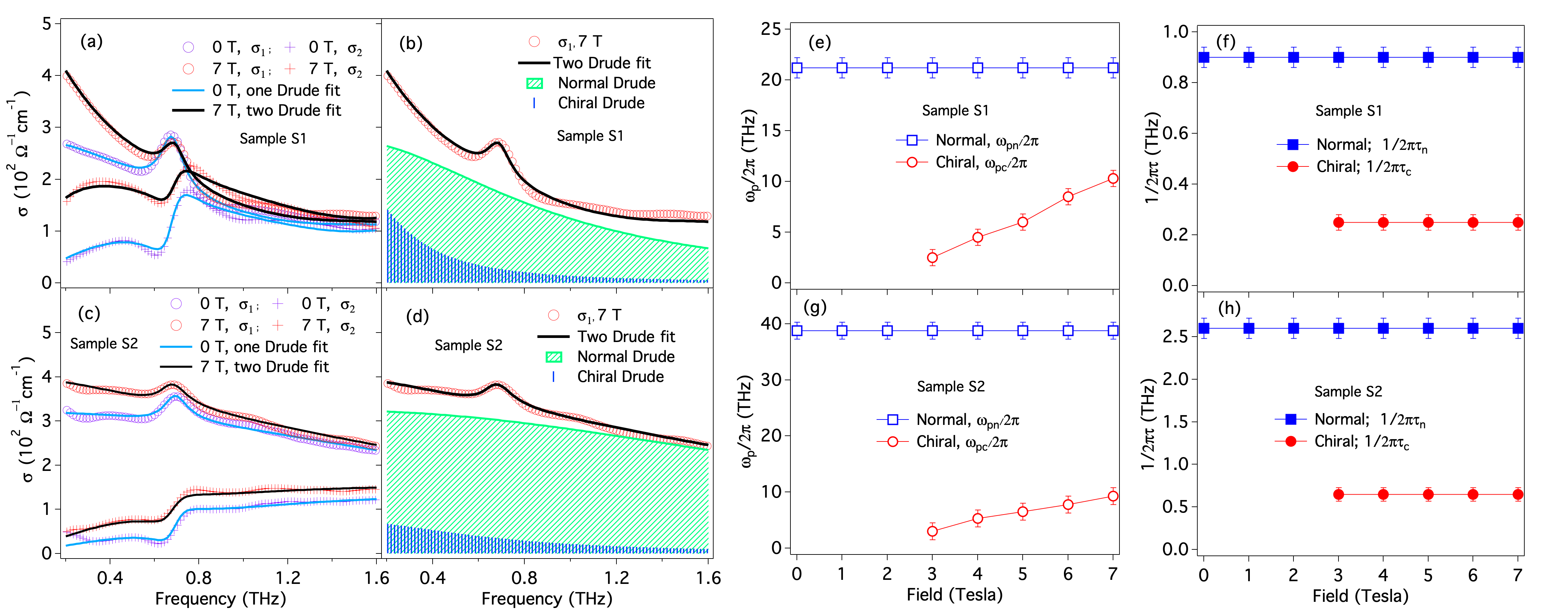}
\caption{ {\bf Dynamical charge pumping and relaxation of the chiral anomaly extracted by Drude-Lorentz fits}. {\bf (a)} and {\bf (b)} Fits to terahertz conductivity of sample S1 with ${\bf E}_{\mathrm{THz}}\parallel {\bf B}$. The sharper Drude oscillator (bule shadowed area) represents the new transport channel from chiral anomaly. {\bf (c)} and {\bf (d)} Fits to terahertz conductivity of sample S2 with ${\bf E}_{\mathrm{THz}}\parallel {\bf B}$. Field dependent Drude plasma frequency in sample S1 {\bf (e)} and in sample S2 {\bf (g)}. The plasma frequencies of chiral transport channel ($\omega_{pc}/2\pi$, Red) directly correspond to chiral charge pumping and are linear functions of field. Scattering rates in sample S1 {\bf (f)} and in sample S2 {\bf (h)}. The chiral scattering rates (1/$2\pi\tau_c$, red) control the dynamical process of chiral anomaly as shown in Figs. \ref{Fig1_new}(d) and in both samples they are much samller than normal bulk scattering rates (1/$2\pi\tau_n$, blue).}
\label{Fig3_new}
\end{figure*}

In Figs. \ref{Fig20}(a) and (b), we show the real part of the THz conductivity at different fields for ${\bf E}_{\mathrm{THz}}\parallel {\bf B}$ and ${\bf E}_{\mathrm{THz}}\perp {\bf B}$ for  Cd$_3$As$_2$ sample S1 (see corresponding $\sigma_{2}$ in the SM) with a low E$_F$ (measured with $\bf{B} \parallel (\overline{1}10)$).  At zero field, $\sigma_{1}$ is characterized by a Drude-like Lorentzian peak with a scattering rate approximately 1 THz, as well as a 0.7 THz phonon \cite{cheng2019large}.  As a field  ${\bf B} \parallel {\bf E}_{\mathrm{THz}} $ is applied, an additional much sharper Drude-like peak rises out of the zero-field $\sigma_{1}$.  That this should be considered a new transport channel with a distinct frequency scale can be seen in that these changes are all at low frequency e.g. at frequencies above 1 THz the data does not change.  To elaborate this point more clearly, we re-plot data of Figs. \ref{Fig20}(a) at each frequency as a function of magnetic field in Figs. \ref{Fig30}(a). One can see in the low frequency region (e.g. below 0.4 THz) the THz conductivity steadily increases with magnetic field with a dependence that is approximately $B^2$ (the expectation for the dc limit of the chiral anomaly). In the high frequency region (above 1 THz) the increasing trend of THz conductivity becomes negligible, indicating the chiral anomaly-related transport is confined to the low frequency region.  $\bf{B} \parallel (\overline{11}2)$ in sample S1 shows a similar effect.  This behavior can be contrasted with ${\bf E}_{\mathrm{THz}}\perp {\bf B}$ that shows a \textit{decrease} in the low frequency conductivity over the \textit{entire} measured spectral range.  For this direction the decreased conductivity is consistent with the usual positive transverse magnetoresistance of conductors in magnetic field. Figs. \ref{Fig20}(a) is the major result of this work.  As shown in  Figs. \ref{Fig20}(c), (d) and \ref{Fig30}(b) another sample S2 measured with $\bf{B} \parallel (\overline{11}2)$ with a larger E$_F$, shows a similar, although more modest effect with scattering rates larger than those of S1. 

To further demonstrate this result, we plot the $\sigma_{1}$ at 0 and 7 T ${\bf E}_{\mathrm{THz}}\parallel {\bf B}$ and their difference for samples S1 and S2 in Figs. \ref{Fig20}(e) and (f) respectively.  One can see that, comparing 0 and 7 T   $\sigma_{1}$ contains an emergent sharper Drude response. The difference in these curves $\Delta\sigma_1=\sigma_{1}$(B)$-\sigma_{1}$(0) at 7 T, reveals a narrower zero-frequency peak that characterizes a new transport channel.    There are also small changes to the phonon that will be discussed elsewhere.

In Figs. \ref{Fig30}(c) and (e), we show the conductivity at 0.3 and 1 THz of sample S1 plotted as a function of field with several different terahertz polarization angles. One can see only the low frequency data (0.3 THz) shows an enhancement of the conductivity, with the effect being most pronounced when ${\bf E} \cdot {\bf B} $ is large.  For crossed ${\bf E} $ and $ {\bf B} $ fields, the magneto-conductivity is negative.   Intermediate angles shows an effect in-between.  Similar data is shown for sample S2 in Figs. \ref{Fig30}(d) and (f).  As shown in the SI Fig. S9 and Fig. S11, the polarization dependence of terahertz conductivity in both samples shows the explicit $\mathrm{cos}^2 \theta$ dependence, consistent with the expectation for the chiral anomaly.

Figs. \ref{Fig3_new}(a) to (d) we show Drude/Drude-Lorentz oscillator fits to the data that characterizes the  ${\bf E}_{\mathrm{THz}}\parallel {\bf B}$ response.  Fit details are given in the Methods section.  In the displayed spectral range, the zero field spectra are fit well by a single Drude feature and Drude-Lorentz oscillator for the 0.7 THz phonon.  The square of the plasma frequency ($\omega_p^2 = Ne^2/ \epsilon m$) is related to the spectral weight of a spectral feature in the conductivity.  At slightly higher frequencies there is an overdamped phonon \cite{cheng2019large,Softphonon2019} which we incorporate through a broad oscillator centered at 1.7 THz.   Small changes in the details of these phonon fits affect none of our main results or conclusions.  As field is increased, the new zero frequency peak emerges and excellent fits can be obtained by the inclusion of an additional sharp Drude term (of width 1/2$\pi\tau_{c}$) with imperceptible changes to the spectral weight and widths of the broader Drude and phonon terms.  The fitting parameters as a function of field are shown in Figs. \ref{Fig3_new}(e) to (h). For both samples, one can see that the plasma frequencies $\omega_{pn}$ and scattering rates 1/2$\pi\tau_{n}$ of the broad Drude oscillator do not have appreciable field evolution.  In contrast, the plasma frequencies $\omega_{pc}$ of the sharper Drude oscillator increases approximately linearly as a function of magnetic field in both samples.  In both samples, 1/2$\pi\tau_{c}$ is approximately one fourth of 1/2$\pi\tau_{n}$.   

\begin{figure*}[t]
\includegraphics[clip,width=6in]{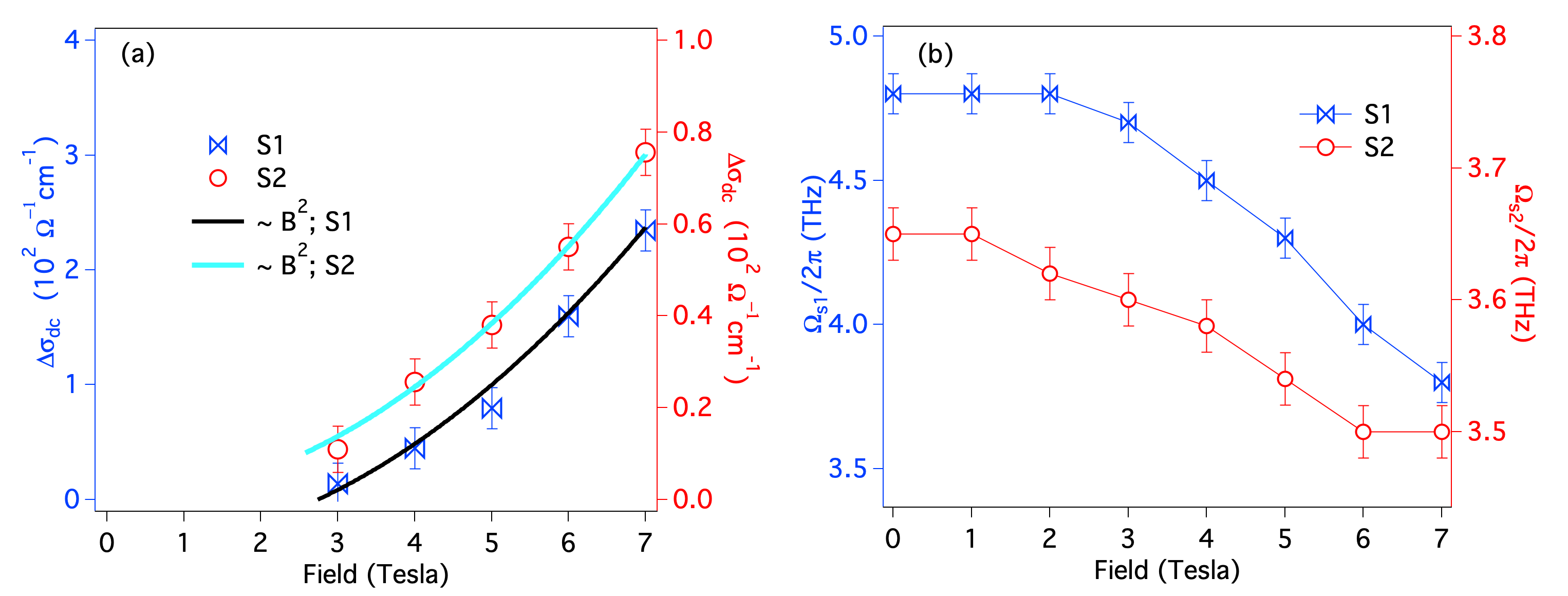}
\caption{ {\bf Intrinsic dc chiral conductivity extrapolated from terahertz conductivity}. {\bf (a)} intrinsic dc magnetoconductivity from chiral anomaly in sample S1 (Blue) and S2 (Red). In both samples, $\Delta\sigma$ follows B$^2$, consistent with the prediction of field dependence of chiral current in semiclassical transport regime. {\bf (b)} phonon oscillator strength in sample S1 (Blue) and S2 (Red). The oscillator strengths in both samples decrease as the chiral conductivity is enhanced by magnetic field.  }
\label{Fig4}
\end{figure*}

This field-induced effect is remarkable and its origin should be carefully considered.   Due to the fact that it is an enhancement of only the low frequency conductivity, it should not be considered due to either a change in the normal scattering rate itself or change in carrier density, but instead the appearance of a parallel transport channel with a {\it new frequency scale}. Moreover the part of the spectra associated with the zero field Drude does not change shows that the effect is not to be associated with effects like spin dependent scattering \cite{van2001temperature} that would manifest as an overall change in scattering rate.  We should also reiterate that an advantage of our contactless THz measurements is that they avoid the artifacts associated with inhomogeneous current paths which occur in very anisotropic conductors when dc currents are applied or in systems with mobility fluctuations \cite{narayanan2015linear,NPA18,TaP_current_jetting_2017,PHE_Na3Bi_PRX_2_2018,parish2003non}.  That the changes to the  spectrum are intrinsic is corroborated by the fact that the phonon's spectral weight [Figs. \ref{Fig4}(b)] decreases as the low frequency conductivity increases (presumably due to increased screening).

In contrast, the appearance of an additional transport channel and new time scale is precisely in agreement with theoretical expectations for the chiral anomaly.   In particular, Burkov showed that with increasing magnetic field an additional Lorentzian-like peak should emerge, the width of which corresponds to the internode scattering rate \cite{burkov2018dynamical}.  Within this picture, the broader Drude represents the normal intraband transitions inside each Weyl cone, whereas the the narrow Drude is evidence for the chiral transport channel, with its spectral weight a direct measure of the chiral population and its width the rate of chiral relaxation 1/2$\pi\tau_{c}$.   The fact that 1/2$\pi\tau_{c}$ shows little field dependence is consistent with predictions.   It is important to point out here that despite the fact that the $( \overline{1}10)$ and $(\overline{11}2)$ fields break the $C_4$ symmetry that protects the Dirac nodes, the chiral anomaly is retained.    For instance discussed in Ref. \cite{cano2017chiral}, a $( \overline{1}10)$ field (as well as other fields that preserve a mirror symmetry) result in a line node that is connected to the other valley via a zeroth Landau level.   For other directions it is important to note that Weyl nodes are generically expected.   For instance, a (001) field separates Dirac nodes into Weyl nodes along the z-axis but does not cause gapping due to the protection by $C_4$.  Then deviations of the field from (001) do not cause gaps to form because Weyl nodes are topological objects \cite{cano2017chiral}.  Moreover, it is important to point out that even with the accepted large g factor of ~16 \cite{narayanan2015linear}, one gets a Zeeman energy of only 3.25 meV.  As this is small as compared to the Fermi energy it does not cause appreciable chiral mixing.  Please see further discussion on this and related points in the SI.

Through simulations of the Drude responses, we can extrapolate the field-induced conductivity $\Delta \sigma_1(B)$ at zero frequency.  $\Delta \sigma_1(B)$ should be equal to the intrinsic dc longitudinal magneto conductivity. In Figs. \ref{Fig4}(a), we show the extrapolated $\Delta \sigma_1(B)$ of both samples. One can see, above 3 T, $\Delta \sigma_1(B)$ shows a typical $B^2$ dependence, which is also consistent with the theoretical prediction of the magnetic field dependence of chiral current in the semiclassical transport regime.

Chiral transport occurs via a build up of the effective chiral electrochemical potential through the balance between chiral pumping and internode scattering.  However, to distinguish a steady-state chiral current, the chiral scattering rate 1/$\tau_{c}$ must be much smaller than 1/$\tau_{n}$.   We find that $1/\tau_{c}$ is approximately one fourth of $1/\tau_{n}$ in both samples.  We can compare this relative size of $1/\tau_{c}$ to $1/\tau_{n}$ in light of prevailing theory.  As mentioned above and shown in Figs. \ref{Fig1_new}(d), there are two potential sources of such scattering.   Charge can undergo large momentum scattering inter-valley scattering ($1/\tau_v$), or can scatter between isospin species at the same valley ($1/\tau_i$).   The larger of these determines $1/\tau_c$.   Their relative scales to $1/\tau_n$ can -- in principle -- be determined from band structure parameters.

As discussed in the SI and in Ref. \cite{Parameswaran14a}, within a Fermi's golden rule approach and for $k_F \ll K_D$ the intervalley scattering rate can be expressed as $\frac{1}{\tau_n}  \Big(  \frac{k_F}{2K_D }  \Big)^4$. There is considerable uncertainty in band structure parameters, but available data \cite{borisenko2014experimental} suggests that $K_D$ is found approximately 1/3 of the way towards $Z$ along the $\Gamma-Z$ line e.g. at approximately $ 0.04  \pm 0.01 \AA^{-1}$ .   Through analysis of the Drude spectral weight and cyclotron mass of S1, one can estimate a Fermi wavevector $k_F$ of 0.0135 $\AA^{-1}$, which gives $1/\tau_v = 0.0008$ $ 1/\tau_n$.   This is much smaller than observed.  The isospin intravalley scattering $ 1/\tau_i$ is suppressed if the band dispersions are completely linear.  However, in a real material, the actual band dispersions will inevitably have quadratic curvature. This quadratic correction mixes chiralities and introduces weak scattering between nodes (See SI for further discussion).  We make an approximation that the impurity potential is spherically symmetric and smooth on the scale of a unit cell and ignore the angular anistropy of the Fermi wavevector and velocity ($v_F$), 1/$\tau_{i}$ can be estimated to be $ \tau^{-1}_n (\frac{ B_3 K_D k_F}{2 v_F})^2$ \cite{Parameswaran14a}, which can be expressed in terms of the cyclotron mass $m^*$ as $1/\tau_i =  \frac{1}{\tau_n}   \Big ( \frac{ B_3 K_D m^*}{ 2\hbar^2} \Big )^2$.  A cyclotron mass of $m^* = 0.03 m_e$ was determined in our previous work \cite{cheng2019large} and an upper bound of $B_3 K_D = 30 \pm 15$ eV $\cdot \AA^2$ can be determined from ARPES data (see SI).   This gives $1/\tau_i = 0.0035$ $ 1/\tau_n$, which as expected for a DSM is larger than $1/\tau_v$, but still much smaller than the $1/\tau_c$  observed.   It may be that impurity potentials that break inversion symmetry and couple chiral sectors at order $k$ need to be considered.  This should be an area of future study.   Recent nonlinear THz experiments have been interpreted in terms of a chiral relaxation rate in the WSM TaAs that is shorter than 1 GHz \cite{levy2020optical}.   Such a slow rate could be due to the much larger separation of nodes in TaAs and/or the lack of isospin scattering.

In this work we have observed an anomalous THz magnetoconductivity effect in the Dirac semimetal Cd$_3$As$_2$.   The effect depends on the relative alignment of the in-plane fields as ${\bf E} \cdot {\bf B}$.   This dependence and the evolution of the functional form of the conductivity is in precise agreement with the theory of the chiral anomaly.   The finite frequency measurements allow the measurement of a new frequency scale that can be associated with the chiral scattering rate.  What is not in precise agreement with prevailing theory is the relative scales of the chiral scattering rate and intranode scattering, as we find the chiral scattering much stronger than predicted.   This raises  questions about the nature of charge transport in the Dirac semimetals and chiral scattering.   It may be that accepted aspects of the band structure need to be revised \cite{roth2018reinvestigating} or that more realistic models of impurity scattering need to be developed.

\footnotesize

\section{Methods }

\textbf{Experimental setup.}  In time-domain terahertz spectroscopy, an 800 nm femtosecond laser pulse is split along two paths and sequentially excites a pair of photoconductive Auston-switch antennae. A broadband THz range pulse is emitted by one antenna, transmitted through the sample under test, and measured at the other antenna by sampling with the split-off pulse.  By varying the length-difference between the two paths, the time dependence of the electric field is measured. The time domain trace is then Fourier transformed into the frequency domain.  Taking the ratio of the transmission through a sample to that of a reference resolves the full complex transmission coefficient. A closed-cycle 7 T superconducting magnet is used to measure field dependent terahertz responses. Samples can be measured with the field direction perpendicular (Voigt geometry) or parallel (Faraday geometry) to the light wavevector \cite{palik1970infrared}. In the current case of thin films deposited on top on an insulating substrate, the transmission can be inverted to obtain the complex conductivity by using the appropriate expression in the thin film approximation \cite{Bing16_1}.

\textbf{Film growth.}  A number of 112 oriented Cd$_3$As$_2$ films were grown by molecular beam epitaxy on (111)B GaAs substrates.   Samples S1 and S2 have thickness of 280 nm and 300 nm respectively.  Further details of the film growth can be found elsewhere \cite{Timo_CdAs_growth_16}. The dc magneto transport of both films is similar to the previous work on the films of this Dirac semimetal.

\textbf{Drude-Lorentz fits.}  To find the scattering rate and spectral weight of the lowest frequency features, the optical conductivity data was fit to a modified Drude/Drude-Lorentz model. We use one Drude oscillator to account for the normal Drude transport of Dirac fermions and the other to account for the chiral Drude response. The expression is:

\begin{equation}
\sigma(\omega)= \epsilon_0 \Big[-{{\omega_{\mathrm{pn}}^2}\over{i \omega - 1/\tau_n}}-{{\omega_{\mathrm{pc}}^2}\over{i \omega - 1/\tau_c}}-{{i\omega\Omega_{s}^2}\over{ \omega_0^2 -\omega^2-i/\tau_0}}-i(\epsilon_{\infty}-1)  \omega \Big].
\label{chik}
\end{equation}

Here, $1/2 \pi \tau_n$ is the intranode (normal) scattering rate of the normal Drude term and $1/2 \pi \tau_c $ is the internode (chiral) scattering rate of the field-induced Drude term. $\omega_{\mathrm{pn}}$ and $\omega_{\mathrm{pc}}$ are their plasma frequencies. $\Omega_{s}$ is the phonon's oscillator strength, $\omega_0$ is the phonon's central frequency, 1/2$\pi\tau_0$ is the phonon linewidth. The background polarizability $\epsilon_{\infty}$ originates from absorptions above the measured spectral range including phonons and interband absorptions.

\textbf{Data availability.} All relevant data are available on reasonable request from N.P.A.

 \bibliography{Quadratic}

 \section{Acknowledgements:  }
 
We would like to thank S. Borisenko, A. Burkov, J. Cano, X. Dai, D. Kharzeev, S. Parameswaran, M. Parish, J. Pixley, and A. Vishwanath for helpful discussions.  Experiments at JHU were supported by the Army Research Office Grant W911NF-15-1-0560.  Work at UCSB was supported by the Vannevar Bush Faculty Fellowship program by the U.S. Department of Defense (grant no. N00014-16-1-2814)
 
 \section{Author contributions:  }
 
B.C. and N.P.A conceived the project. B.C. performed and analyzed the magneto-TDTS measurements. Films were developed by T.S. and S.S.   T.S. performed dc transport measurements.  The manuscript was written by B.C. and N.P.A. with input from all authors.

 \section{Additional information:}

\textbf{Competing financial interests:} The authors declare no competing financial interests.

\normalsize

\newpage

\setcounter{figure}{0}
\setcounter{equation}{0}
\setcounter{section}{0}
\begin{widetext}

\section{1. I\lowercase{maginary parts of optical conductivity with} ${\bf E}_{\mathrm{THz}}\parallel {\bf B}$ \lowercase{and} ${\bf E}_{\mathrm{THz}}\perp {\bf B}$ \lowercase{at} 6 K}

\renewcommand{\thefigure}{S\arabic{figure}}

In the main text, we emphasized the real parts of the optical magneto-conductivity $\sigma_1$ with ${\bf E}_{\mathrm{THz}}\parallel {\bf B}$ and demonstrated that the application of $  {\bf E} \cdot {\bf B} $ introduces a new transport channel to Cd$_3$As$_2$.  We show in Fig. \ref{imag_cond}, the imaginary parts of optical conductivity $\sigma_2$ with ${\bf E}_{\mathrm{THz}}\parallel {\bf B}$ and ${\bf E}_{\mathrm{THz}}\perp {\bf B}$. As shown in Figs. \ref{imag_cond}(a) and (c), when increasing magnetic field, $\sigma_2$ is gradually enhanced. This is consistent with our discussion of magnetic field dependence of $\sigma_1$ that a new transport channel appears after applying $  {\bf E} \cdot {\bf B} $. In contrast, when ${\bf E}_{\mathrm{THz}}\perp {\bf B}$, as shown in Figs. \ref{imag_cond}(b) and (d), both samples' $\sigma_2$ are gradually suppressed by magnetic field.   This is consistent with the usual positive magneto resistivity that occurs if the current is perpendicular to magnetic field.

\begin{figure*}[htp]
\includegraphics[clip,width=7in]{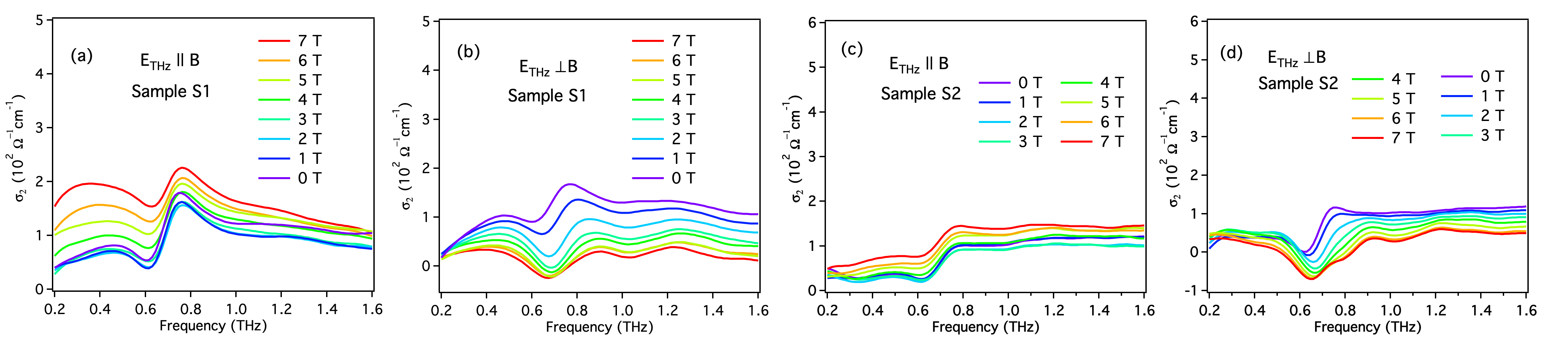}
\caption{Imaginary parts of optical conductivity of Sample S1 with (a) ${\bf E}_{\mathrm{THz}}\parallel {\bf B}$ and (b) ${\bf E}_{\mathrm{THz}}\perp {\bf B}$. Imaginary parts of optical conductivity of Sample S2 with (c) ${\bf E}_{\mathrm{THz}}\parallel {\bf B}$ and (d) ${\bf E}_{\mathrm{THz}}\perp {\bf B}$.}
\label{imag_cond}
\end{figure*}

\section{2. T\lowercase{wo} D\lowercase{rude fit to }TH\lowercase{z conductivity at high magnetic field}   }

As shown in main text Fig. 2, when ${\bf E}_{\mathrm{THz}}\parallel {\bf B}$, the magnetic field enhances the THz conductivity in the low frequency region but does not introduce notable changes above 1 THz. Just through inspecting the THz conductivity data themselves, we could conclude magnetic field induces a new transport channel that can be described by an emergent and sharper Drude oscillator. At zero field, the free charge transport part of THz conductivity can be well described by one Drude oscillator. At high field, we need to add a new Drude oscillator to describe the emergent and sharper Drude. To elaborate our points more clearly, we show two Drude and one Drude fit to the THz conductivity of Sample S1 at 7 T Fig. \ref{Two_va_one}(a) and (b) respectively. One can see the two Drude fit captures well all features of the real and imaginary parts of THz conductivity. In contrast, because of the emergence of a sharp low frequency THz conductivity component, no matter how one tunes the Drude fitting parameters, a single Drude oscillator cannot reproduce the low frequency THz conductivity. 

We show detailed fitting examples of zero field and 7 T terahertz conductivity (${\bf E}_{\mathrm{THz}}\parallel {\bf B}$) of Sample S1 in Fig. \ref{Two_va_one_detail}(a) and (b) respectively. Responses from electron transport and phonons could be clearly isolated in our terahertz conductivity data. The phonon mode could be well described by a Lorentzian oscillator near 0.7 THz. Please note the phonon mode at 0.7 THz has weak field dependence. Because the phonon mode is relatively sharp, its field dependence could be exactly captured by fitting with a Lorentzian oscillator. The spectral weight of this phonon mode decreases (Fig. 5(b) in main text) with enhancing magnetic field. We have discussed its origin in main text. In the electronic part, from the Fig. 2 and Fig. 3 in main text, one could see the terahertz conductivity (${\bf E}_{\mathrm{THz}}\parallel {\bf B}$) above 1 THz (the tail of the Drude response) will not evolve with magnetic field. This means the terahertz conductivity should include a field independent Drude response in all fields. We use the green curves [Fig. \ref{Two_va_one_detail}(a) and (b)] to represent this field independent part. It includes a Drude term and a weak and broad oscillator centered at 1.7 THz. Comparing to the zero-field conductivity, the primary effect of ${\bf E}_{\mathrm{THz}}\parallel {\bf B}$ is the emergent Drude response [the blue curve in Fig. \ref{Two_va_one_detail}(b)]. This field induced Drude is apparently much sharper than the field independent Drude. As we discussed in main text, we attribute it to the chiral anomaly. To fit the terahertz conductivity in all fields, we keep the field-independent part unchanged (green curves), and use a field dependent Lorentzian oscillator (magenta curves) near 0.7 THz, and a sharper Drude term (blue curves) to capture all features of the data. The black curves represent the full fits and they are well reproduced all data.

As discussed in the main text, Sample S2 shows a weaker version of the same effects with applied field. Although the disagreement is not as pronounced as for Sample S1, a one Drude fit is also inadequate for Sample S2. We show the two Drude and one Drude fits to the THz conductivity of Sample S2 at 7 T in  Fig. \ref{Two_va_one}(c) and (d) respectively. Similar to the case of Sample S1, the two Drude fit well captures all features of the real and imaginary parts of THz conductivity of Sample S2. By using one Drude fit, as shown in  Fig. \ref{Two_va_one}(d), the fit captures the real part of THz conductivity but does not well reproduce the imaginary part comparing to two Drude fit. In Fig. \ref{Two_va_one_detail}(c) and (d), we show the detailed fitting examples of zero field and 7 T terahertz conductivity (${\bf E}_{\mathrm{THz}}\parallel {\bf B}$) of Sample S2 respectively. The fitting process is the same as we fit data of Sample S1. In the course of our THz study of Cd$_3$As$_2$, we have taken a large amount of with different samples, temperature dependence and polarization dependence. These are shown in the main text and supplementary material. Furthermore, our time-domain THz spectrometer can determine the real and imaginary parts of THz conductivity simultaneously. We want to emphasize that the fitting and analysis of our data have been constrained by all these factors.

\clearpage

\begin{figure*}[t]
\includegraphics[clip,width=5in]{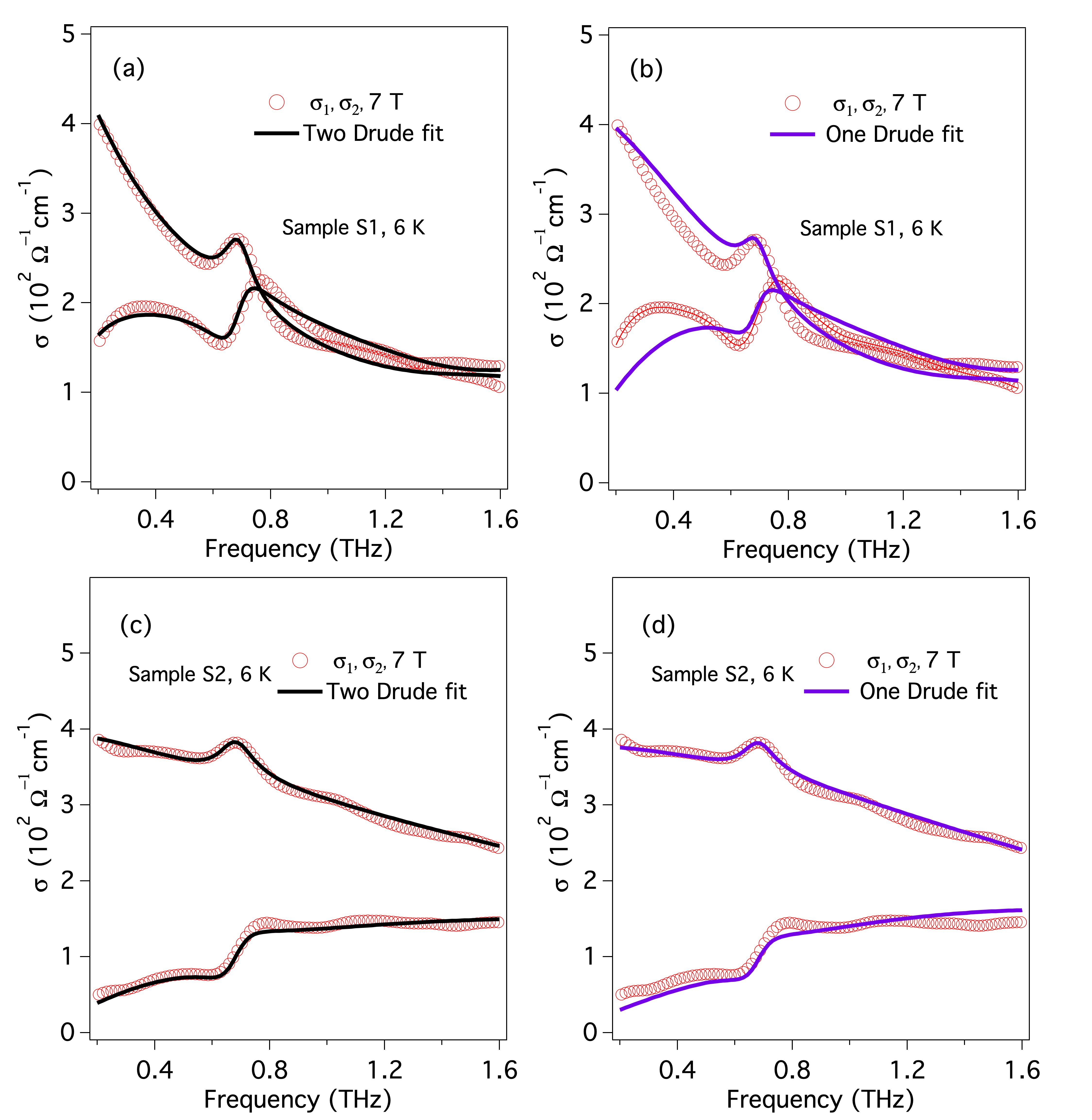}
\caption{Two Drude fit (a) and one Drude fit (b) to the THz conductivity of Sample S1 with ${\bf E}_{\mathrm{THz}}\parallel {\bf B}$. Two Drude fit (c) and one Drude fit (d) to the THz conductivity of Sample S2 with ${\bf E}_{\mathrm{THz}}\parallel {\bf B}$.}
\label{Two_va_one}
\end{figure*}

\begin{figure*}[t]
\includegraphics[clip,width=5in]{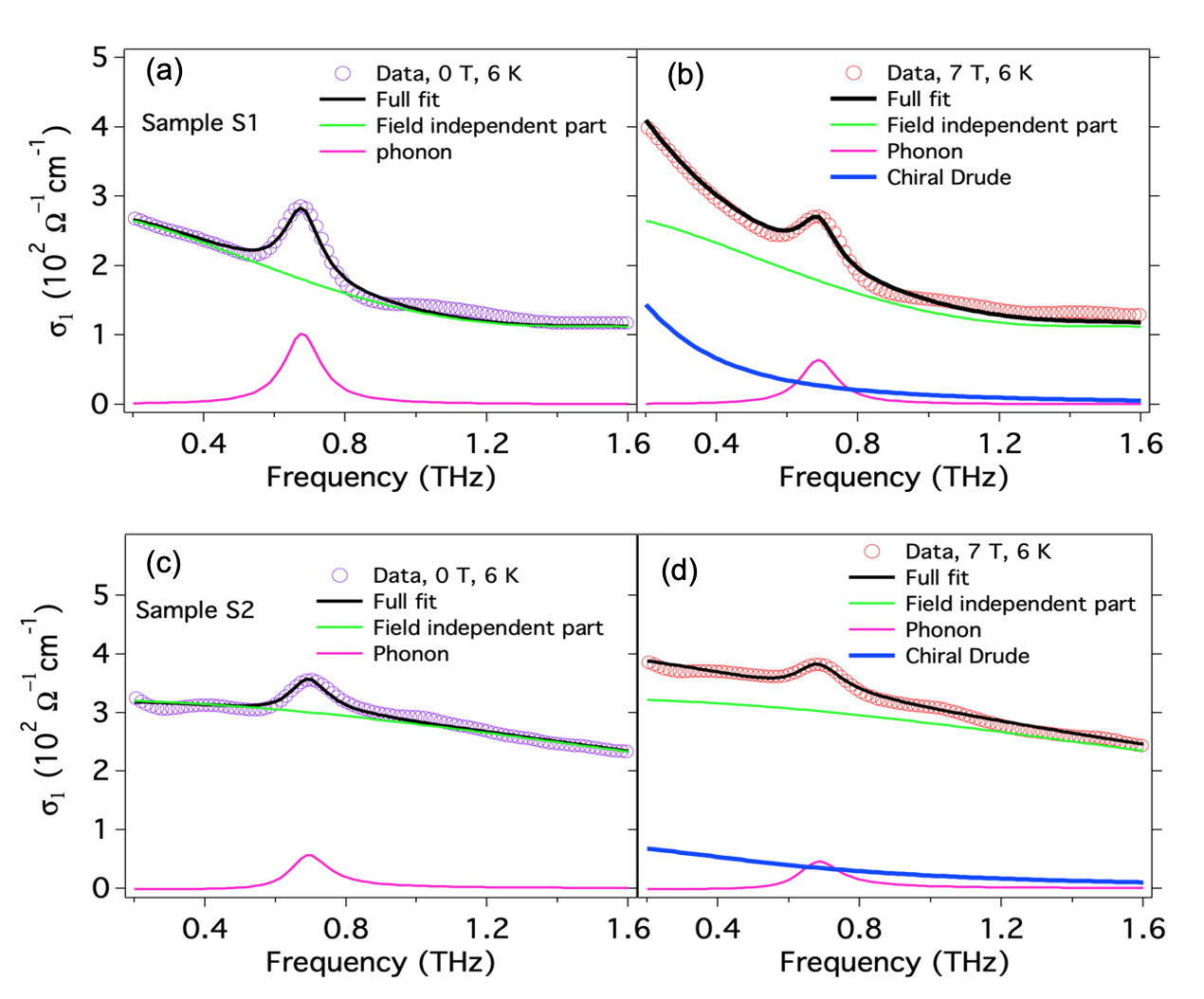}
\caption{Drude-Lorentz fitting details of terahertz conductivity at 0 T (a) and 7 T (b) of Sample S1 with ${\bf E}_{\mathrm{THz}}\parallel {\bf B}$. Drude-Lorentz fitting details of terahertz conductivity at 0 T (c) and 7 T (d) of Sample S2 with ${\bf E}_{\mathrm{THz}}\parallel {\bf B}$.}
\label{Two_va_one_detail}
\end{figure*}

\clearpage

\section{3. M\lowercase{agneto }TH\lowercase{z conductivity at 100} K \lowercase{and} 150 K  } 

To further explore the chiral magnetic effect, we also measured the THz magneto-conductivity of Samples S1 and S2 at 100 K and 150 K. In Fig. \ref{cond_S1_100K}(a), we show the real part of field dependent THz conductivity of Sample S1 at 100 K. One can see the field dependence at 100 K is very similar to the data at 6 K (main text Fig. 2). Above 0.8 THz, $\sigma_1$ has little field dependence, but below 0.5 THz a sharp Drude response gradually grows with increasing magnetic field. The field-enhanced THz conductivity itself is the positive longitudinal magneto-conductivity, a sign of the chiral magnetic effect. We again used the Drude-Lorentz model to fit the data. As shown in Fig. \ref{cond_S1_100K}(d) the THz conductivity at zero field could be well described by one broad electronic Drude term and a sharp phonon Lorentzian term. Above 3 T, without changing anything of the zero field Drude terms, we can use an extra narrow Drude term to describe the field induced THz conductivity. The fitting examples are shown in Fig. \ref{cond_S1_100K}(d) to (i). All fitting parameters are shown in Fig. \ref{cond_S1_100K}(b) and 2(c). One can see the field dependences of plasma frequency and scattering rate of normal and chiral transport are similar to that of 6 K (main text Fig. 3). We show the field-dependent THz conductivity of Sample S1 at 150 K in Fig. \ref{cond_S1_150K}, the field-dependent THz conductivity of Sample S2 at 100 K and 150 K in Fig. \ref{cond_S2_100K} and Fig. \ref{cond_S2_150K} respectively. The results are similar to the data of 100 K of Sample S1.

\begin{figure*}[b]
\includegraphics[clip,width=7.2in]{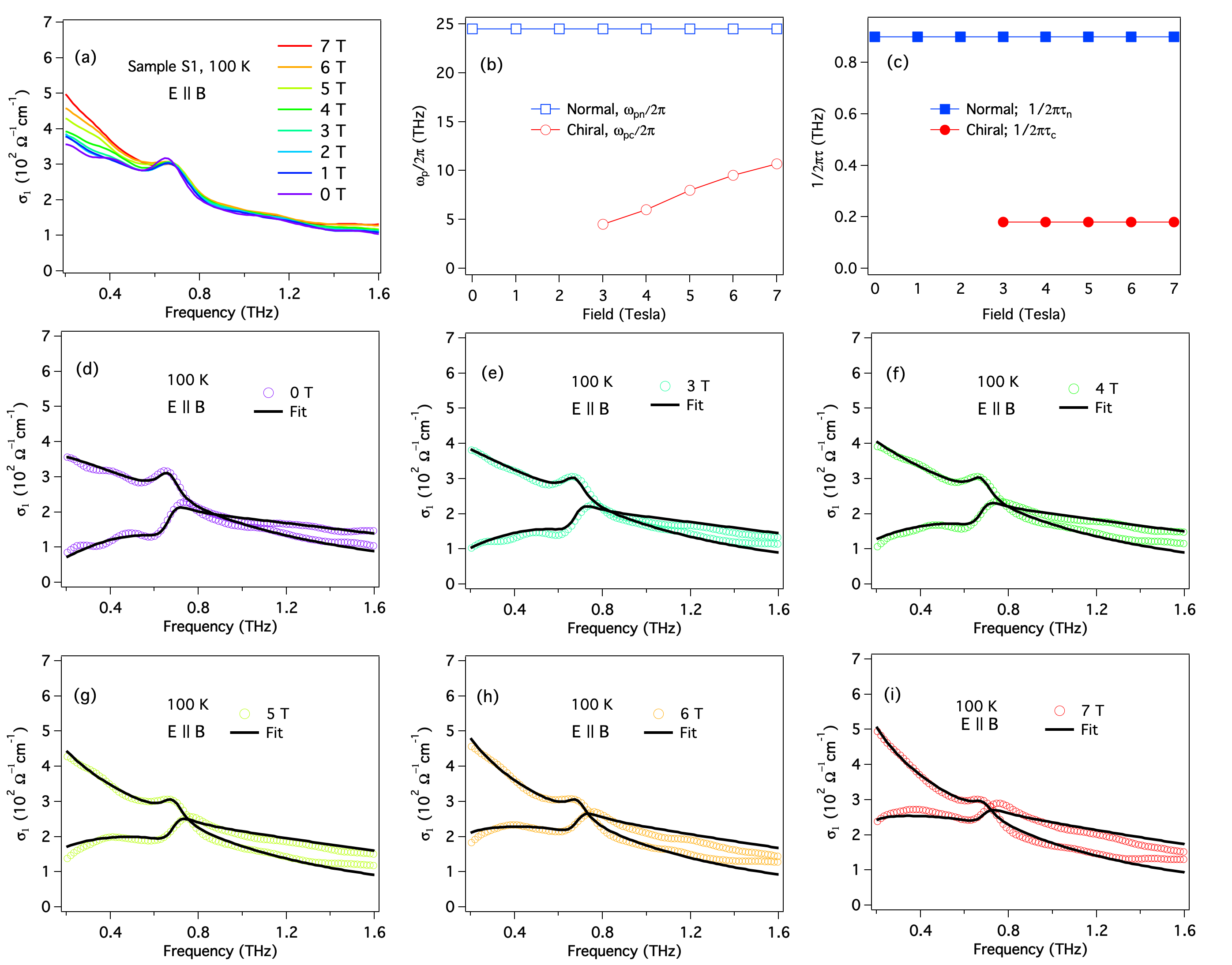}
\caption{(a) THz conductivity at different magnetic fields in the geometry of ${\bf E}_{\mathrm{THz}}\parallel {\bf B}$ with $\bf{B} \parallel (\overline{1}10)$ for sample S1. The data are taken at 100 K. (b) Field dependent Drude plasma frequencies of the normal transport channel and the chiral transport channel. (c) Field dependent scattering rates of the normal transport channel and the chiral transport channel at 100 K. Drude-Lorentz fits for complex THz conductivity at (d) 0 T, (e) 3 T, (f) 4 T, (g) 5 T, (h) 6 T, and (i) 7 T. }
\label{cond_S1_100K}
\end{figure*}

\clearpage

\begin{figure*}[htp]
\includegraphics[clip,width=7.2in]{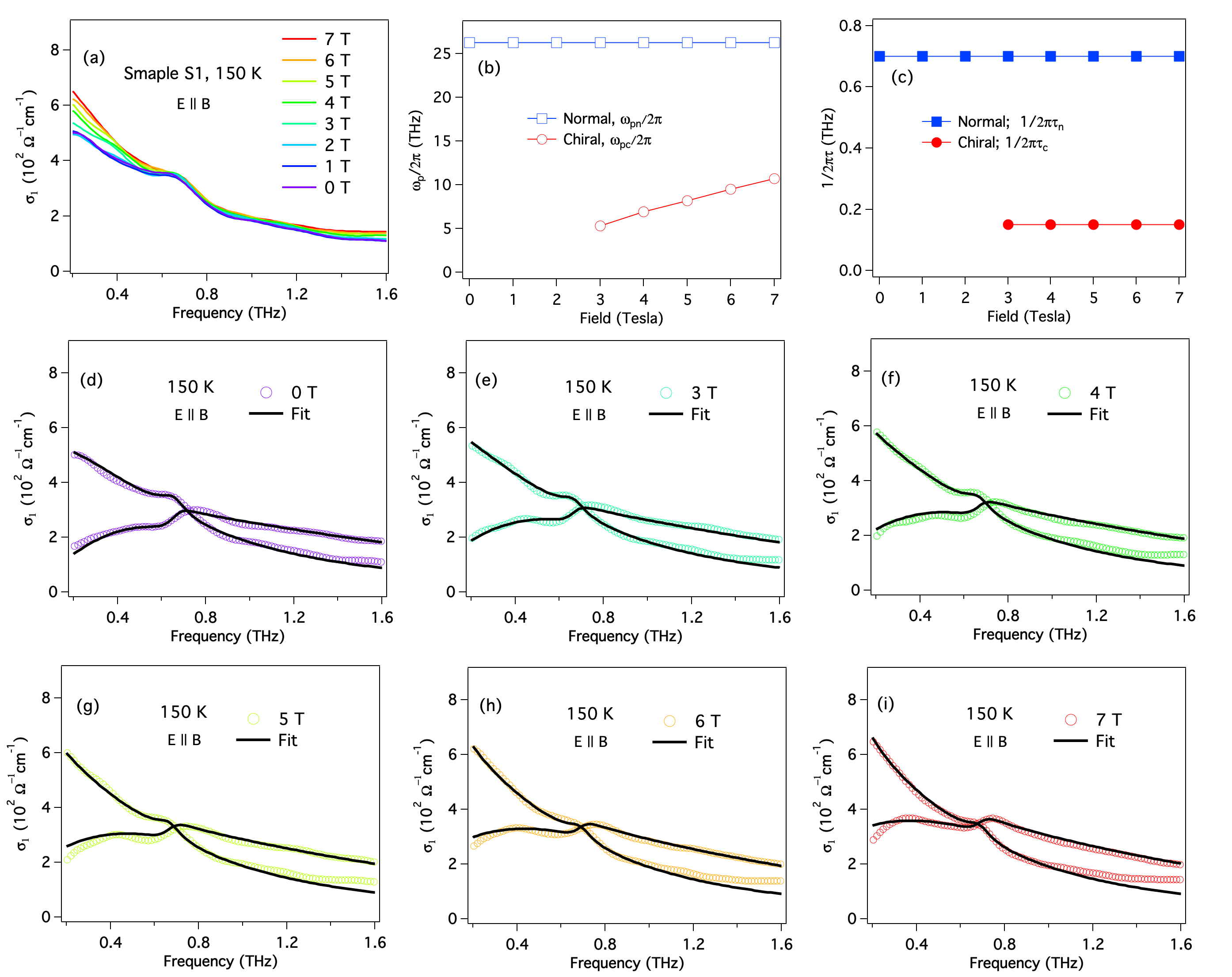}
\caption{(a) THz conductivity at different magnetic fields in the geometry of ${\bf E}_{\mathrm{THz}}\parallel {\bf B}$ with $\bf{B} \parallel (\overline{1}10)$ for sample S1. The data are taken at 150 K. (b) Field dependent Drude plasma frequencies of the normal transport channel and the chiral transport channel. (c) Field dependent scattering rates of the normal transport channel and the chiral transport channel at 150 K. Drude-Lorentz fits for complex THz conductivity at (d) 0 T, (e) 3 T, (f) 4 T, (g) 5 T, (h) 6 T, and (i) 7 T.}
\label{cond_S1_150K}
\end{figure*}

\clearpage

\begin{figure*}[htp]
\includegraphics[clip,width=7.2in]{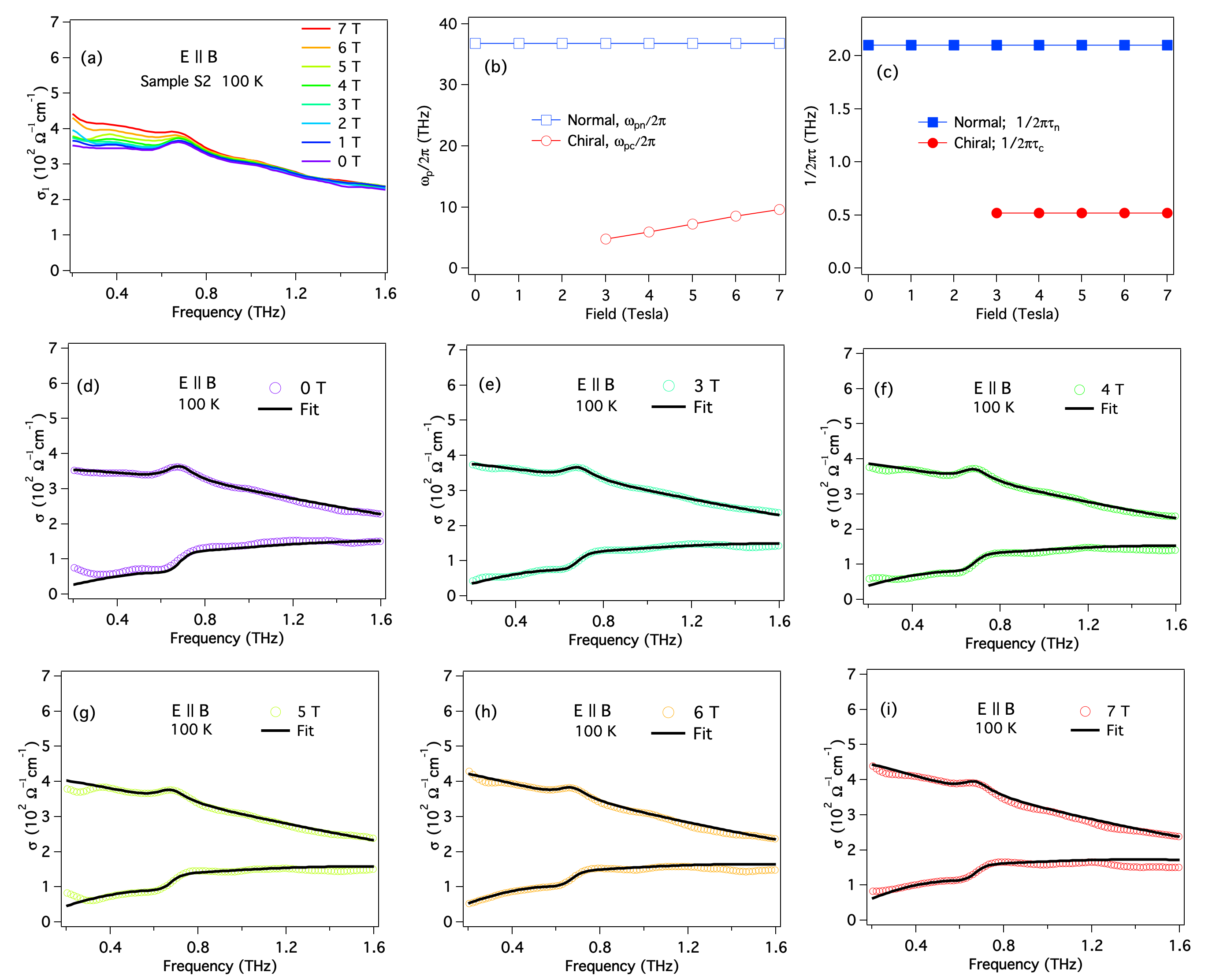}
\caption{(a) THz conductivity at different magnetic fields in the geometry of ${\bf E}_{\mathrm{THz}}\parallel {\bf B}$ with $\bf{B} \parallel (\overline{11}2)$ for sample S2. The data are taken at 100 K. (b) Field dependent Drude plasma frequencies of the normal transport channel and the chiral transport channel at 100 K. (c) Field dependent scattering rates of the normal transport channel and the chiral transport channel. Drude-Lorentz fits for complex THz conductivity at (d) 0 T, (e) 3 T, (f) 4 T, (g) 5 T, (h) 6 T, and (i) 7 T.}
\label{cond_S2_100K}  
\end{figure*}

\clearpage

\begin{figure*}[htp]
\includegraphics[clip,width=7.2in]{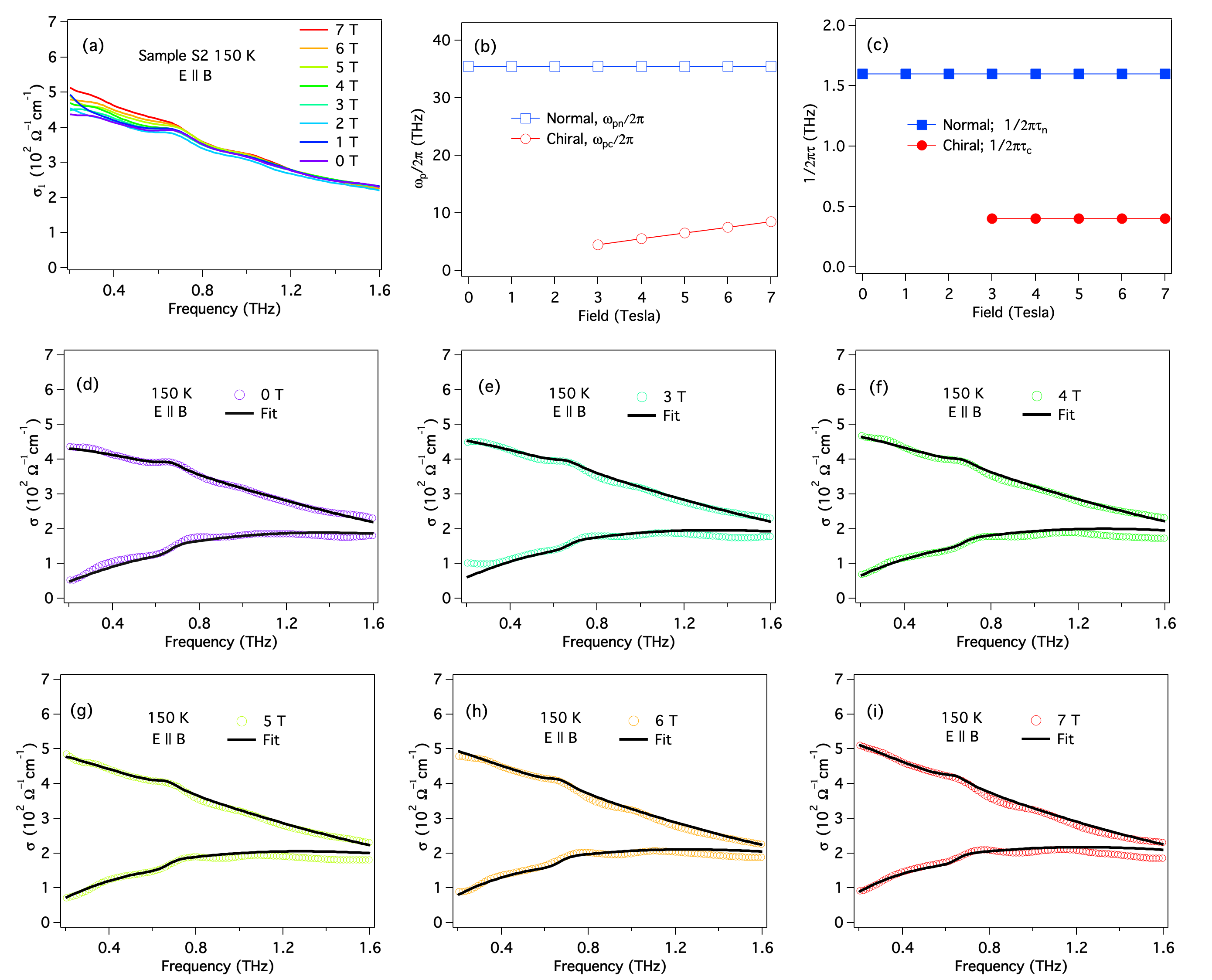}
\caption{(a) THz conductivity at different magnetic fields in the geometry of ${\bf E}_{\mathrm{THz}}\parallel {\bf B}$ with $\bf{B} \parallel (\overline{11}2)$ for sample S2. The data are taken at 150 K. (b) Field dependent Drude plasma frequencies of the normal transport channel and the chiral transport channel at 150 K. (c) Field dependent scattering rates of the normal transport channel and the chiral transport channel. Drude-Lorentz fits for complex THz conductivity at (d) 0 T, (e) 3 T, (f) 4 T, (g) 5 T, (h) 6 T, and (i) 7 T.}
\label{cond_S2_150K}
\end{figure*}

\clearpage

\section{4. P\lowercase{olarization dependent }TH\lowercase{z magneto-conductivity at 6} K }

Angular magneto-conductivity and planar Hall effect have been extensively used to study the conductivity anisotropy in magnetic materials. Here, we provide polarization-dependent THz magneto-conductivity of the Dirac semimetal Cd$_3$As$_2$. All polarization-dependent data shown here are taken at 6 K. Our THz polarization-dependent measurement is similar to dc angular magneto-conductivity and planar Hall effect measurements but provides extra information at finite frequency. During the measurement, we fixed the direction of magnetic field to be horizontal, in-plane, and along $(\overline{1}10)$ or $(\overline{11}2)$ of the samples.  Then the in-plane magnetic field {\bf B} defines two special orthogonal directions {\bf a} and {\bf b} ({\bf a} $\parallel$ {\bf B} and {\bf b} $\perp$ {\bf B}). As shown in Fig. 2 of main text, Cd$_3$As$_2$ films show chiral anomaly-related positive longitudinal magneto-conductivity when {\bf E} $\parallel$ {\bf a}. We label the THz conductivity in the geometry of {\bf E} $\parallel$ {\bf a} $\parallel$ {\bf B} as $\sigma_{\parallel}$. In contrast, Cd$_3$As$_2$ films show large negative longitudinal magneto-conductivity when {\bf E} $\parallel$ {\bf b} $\perp$ {\bf B}.   The conductivity in this geometry we label $\sigma_{\perp}$.

We now allow the THz electric field to be applied in an arbitrary direction with respect to the applied magnetic field.    By using polarizers in front of the sample, we can vary the linear polarization of THz pulse.  We label the reference frame set by the polarization of the THz electric field as ${\bf x'}- {\bf y'}$ with ${\bf E}  \parallel {\bf x'}$ and ${\bf E}  \perp {\bf y'}$.  And therefore we label the conductivity measured along and perpendicular to the polarization direction of THz electric field as $\sigma_{xx}$ and $\sigma_{xy}$ respectively. Fig. \ref{SMcartoon} depicts the details of the geometry and reference frames.    The conductivities are related as

\begin{equation}
\sigma_{xx}(\omega)= \sigma_{\perp}\mathrm{sin}^2\theta  + \sigma_{\parallel}\mathrm{cos}^2\theta =  \sigma_{\perp}+\Delta\sigma \mathrm{cos}^2\theta;\quad \sigma_{xy}(\omega) = \Delta\sigma  \mathrm{sin}\theta \mathrm{cos}\theta.
\label{angle_cond_simulation}
\end{equation}
 \noindent where $\Delta\sigma$ =  $\sigma_{\parallel}$ -- $\sigma_{\perp}$ determines the anisotropy in conductivity induced by magnetic field. $\theta$ is the linear polarization angle of THz electric field referenced to the ${\bf B}$ direction as shown in the schematic in Fig. \ref{SMcartoon}. In dc transport measurements, people also refer $\theta$ to the field tilt angle between dc electric field and and applied magnetic field\cite{chiral_2016}.

In Fig. \ref{SI_angle_dep_s1}, we show the THz magneto-conductivity $\sigma_{xx}$ of Sample S1 at $-$90$^\circ$, $-$75$^\circ$, $-$60$^\circ$, $-$45$^\circ$, $-$30$^\circ$, $-$15$^\circ$, 0$^\circ$, 15$^\circ$, 30$^\circ$, 45$^\circ$, 60$^\circ$ and 75$^\circ$. One can see at $\theta$ = 0$^\circ$, $\sigma_1$ shows the same positive longitudinal magneto-conductivity with increasing magnetic field as we show in main text Fig. 2. With increasing or decreasing the linear polarization angle, the positive longitudinal magneto-conductivity is gradually suppressed and finally not visible at $\pm$45$^\circ$. At $\theta$ = $-$90$^\circ$, $\sigma_{xx}$ displays a conventional negative longitudinal magneto-conductivity. These results are consistent with a chiral anomaly-related response as claimed by previous dc transport works\cite{chiral_2016,TaAs_chiral_2015,ZrTe5_chiral_2015,PHE_Na3Bi_PRX_2_2018}.

To see this point more clearly, we plot the 7 T $\sigma_{xx}$ at several polarization angles in Fig. \ref{SI_angle_fit_s1}(a). As we showed in Eq. \ref{angle_cond_simulation}, these polarization dependent $\sigma_{xx}$ are not independent. $\sigma_{xx}$ at each polarization angle is just determined by $\sigma_{\parallel}$, $\sigma_{\perp}$ and polarization angle $\theta$. In Fig. \ref{SI_angle_fit_s1}(a), we set THz conductivity at 0$^\circ$ as $\sigma_{\parallel}$ and at 90$^\circ$ as $\sigma_{\perp}$. The black curves are the fitting by using Eq. \ref{angle_cond_simulation} to THz conductivity measured at different terahertz polarization angles. One can see the fittings capture all features and reproduce THz conductivities at each polarization angle. Simultaneously, we plot THz conductivity at 0.25 THz and 0.4 THz as a function of polarization angle $\theta$ in Fig. \ref{SI_angle_fit_s1}(b). One can see THz conductivity at both frequencies could be well described by $\sigma_{\perp}+\Delta\sigma \mathrm{cos}^2\theta$ (Eq. \ref{angle_cond_simulation}). From these data and simulations, one can see although we vary the polarization of THz electric field, $\sigma_{\parallel}$ and $\sigma_{\perp}$ are the same. They simultaneously contribute to the polarization-dependent conductivity $\sigma_{xx}$ by multiplying their own anisotropic factors $\mathrm{sin}^2$$\theta$ and $\mathrm{cos}^2$$\theta$ respectively. These results support our inference that the positive longitudinal magneto-conductivity observed with {\bf E} $\parallel$ {\bf B} is intrinsic and comes from the chiral anomaly. In Fig. \ref{SI_angle_dep_s2} and \ref{SI_angle_fit_s2}, we show the polarization dependent THz magneto-conductivity and some fittings of Sample S2. The field and polarization dependencies of THz conductivity, and our main conclusion of this sample are very similar to the results of Sample S1.
 
To investigate the field- and polarization-dependent behaviors of planar Hall conductivity $\sigma_{xy}$,  we further performed the rotation angle measurements with different linear polarizations of THz electric field. Light polarization measurement such as Faraday and Kerr rotation is effectively equivalent to measuring Hall conductivity\cite{Mn3Sn_Kerr_2018,LiangWu_phonon_2015,TCI_Bing_PRL}. The details for how we perform the rotation angle measurement by THz spectroscopy could be found in our previous publication \cite{wu2016quantized}. Please note that here we apply an in-plane magnetic field rather than an out-of-plane magnetic field. The measurement geometry between the sample, the THz polarization and magnetic field is shown in Fig. \ref{SMcartoon}. In the Faraday geometry (with out-of-plane magnetic field), measuring the rotation angle $\varphi$ is equivalent to measuring the Hall conductivity $\sigma_{xy}$\cite{LiangWu_phonon_2015,TCI_Bing_PRL}. However, in Voigt geometry (with in-plane magnetic field), although one still can measure a quantity called planar Hall conductivity $\sigma_{xy}$, this effective Hall conductivity is a way of quantifying the anisotropy of the conductivity as per Eq. \ref{angle_cond_simulation}.   It does not satisfy the Onsager relation for a true Hall response\cite{PHE_Na3Bi_PRX_2_2018}.  The rotation of THz polarization in Voigt geometry linearly scales with the transverse electric field induced by in-plane magnetic field e.g. $\varphi$ = $k$$\sigma_{xy}$, where $k$ is a field dependent quantity but will be a constant under a fixed magnetic field. The planar Hall conductivity induced by chiral anomaly-related transport will follow: $\sigma_{xy}$ = $\Delta\sigma$sin$\theta$cos$\theta$\cite{PHE_chiral_theory_PRL_2017,PHE_chiral_theory_PRB_2017}. Here, $\Delta\sigma$ =  $\sigma_{\parallel}$ -- $\sigma_{\perp}$ and $\theta$ is the polarization angle of the THz pulse. When $\theta$ = 0$^\circ$ and $\pm$90$^\circ$, the rotation angle $\varphi$ = $k$$\sigma_{xy}$ = 0 even if the in-plane magnetic field is applied. One can see the rotation angle data in Fig. \ref{SI_rotation_s1} and Fig. \ref{SI_rotation_s2}. In both samples, the rotation angle $\varphi$ are nearly zero and do NOT have field evolution when $\theta$ = 0$^\circ$ and $\pm$90$^\circ$. In contrast, the rotation angle $\varphi$ has clear field dependence when $\theta$ is away from 0$^\circ$ and 90$^\circ$. To look at the THz polarization dependent behavior of planar Hall conductivty $\sigma_{xy}$ in the maximum magnetic field magnitude 7 Tesla, we further plot the real rotation angle at 0.3 THz of Sample S1 and S2 as a function of THz polarization angle $\theta$ in Fig. \ref{PHE_S1} and Fig. \ref{PHE_S2} respectively. A sin$\theta$cos$\theta$ fit can very nicely reproduce the polarization dependent rotation angle (planar Hall conductivity $\sigma_{xy}$). We want to point out that when the THz polarization angle $\theta$ = 0$^\circ$ ({\bf E} $\parallel$ {\bf B}) and $\theta$ = 90$^\circ$ ({\bf E} $\perp$ {\bf B}), the rotation angle (the planar Hall conductivity $\sigma_{xy}$) is very close to zero which can also be clearly seen in Fig. \ref{PHE_S1} and Fig. \ref{PHE_S2}. These features further support our measurements of THz conductivity with {\bf E} $\parallel$ {\bf B} and {\bf E} $\perp$ {\bf B} are intrinsic.

\begin{figure*}[htp]
\includegraphics[clip,width=2.5in]{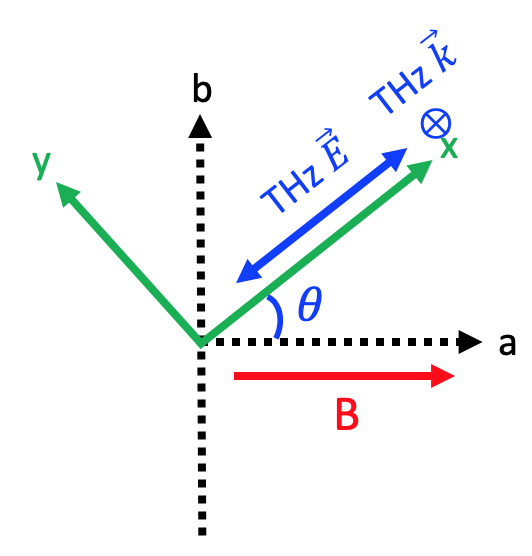}
\caption{The dashed frame represents the coordinate fixed with samples. Its {\bf a} axis is parallel to magnetic field. The {\bf x}$-$ {\bf y} coordinate is rotating along with the polariztion of THz electric field. the {\bf x} axis is parallel to THz electric field. The angle between the polarization of THz electric field and magnetic field is labelled as $\theta$. In our measurements, $\theta$ varies from $-$90$^\circ$ to 90$^\circ$. }
\label{SMcartoon}
\end{figure*}

\clearpage
 
\begin{figure*}[htp]
\includegraphics[clip,width=6.5in]{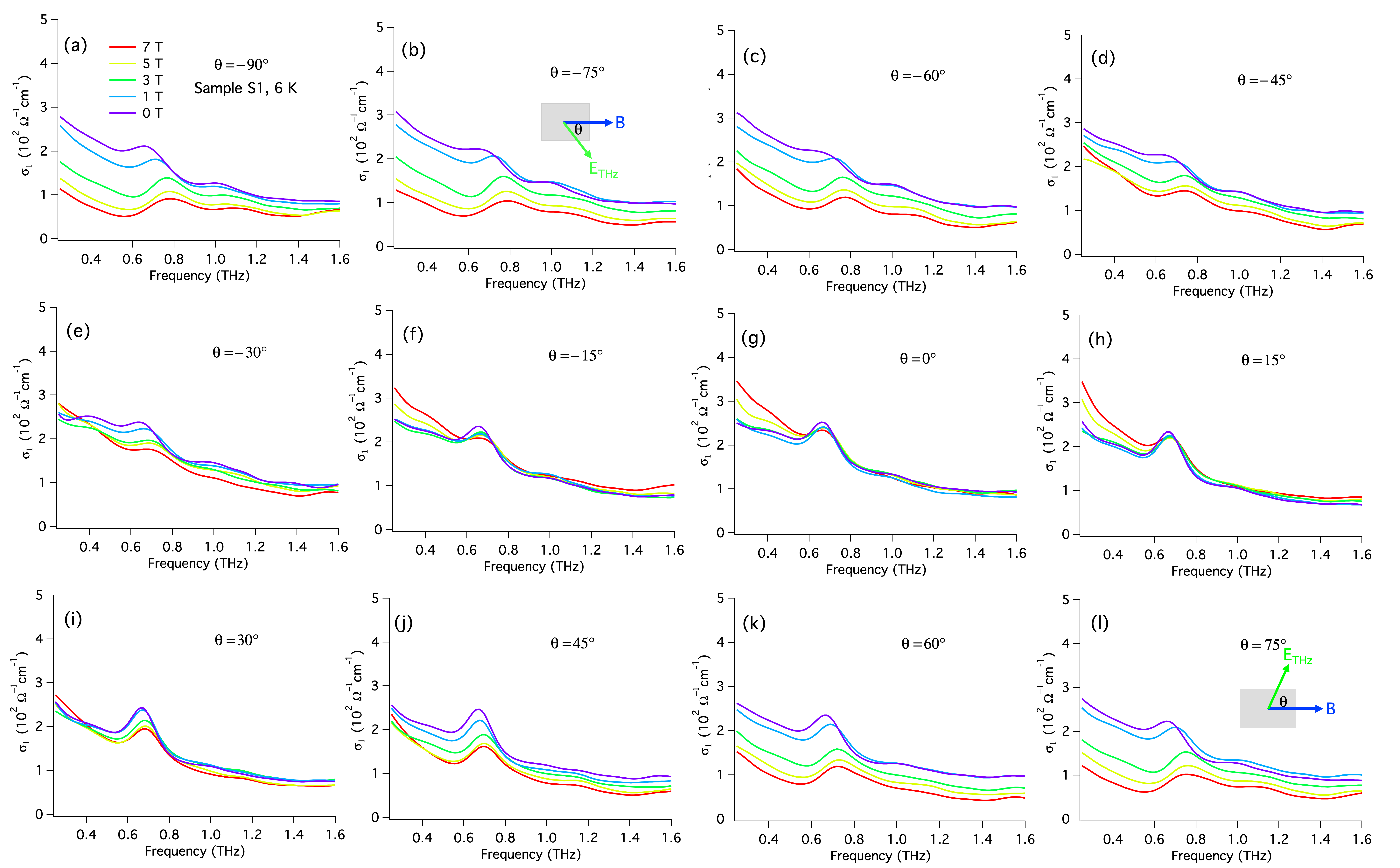}
\caption{Longitudinal THz magneto-conductivity of Sample S1 at the polarization angle of (a) $-$90$^\circ$, (b) $-$75$^\circ$, (c) $-$60$^\circ$, (d) $-$45$^\circ$, (e) $-$30$^\circ$, (f) $-$15$^\circ$, (g) 0$^\circ$, (h) 15$^\circ$, (i) 30$^\circ$, (j) 45 $^\circ$, (k) 60$^\circ$, (l) 75$^\circ$. The data are taken at 6 K. {\bf B} is parallel to ($\overline{1}$10). }
\label{SI_angle_dep_s1}
\end{figure*}

\begin{figure*}[htp]
\includegraphics[clip,width=6in]{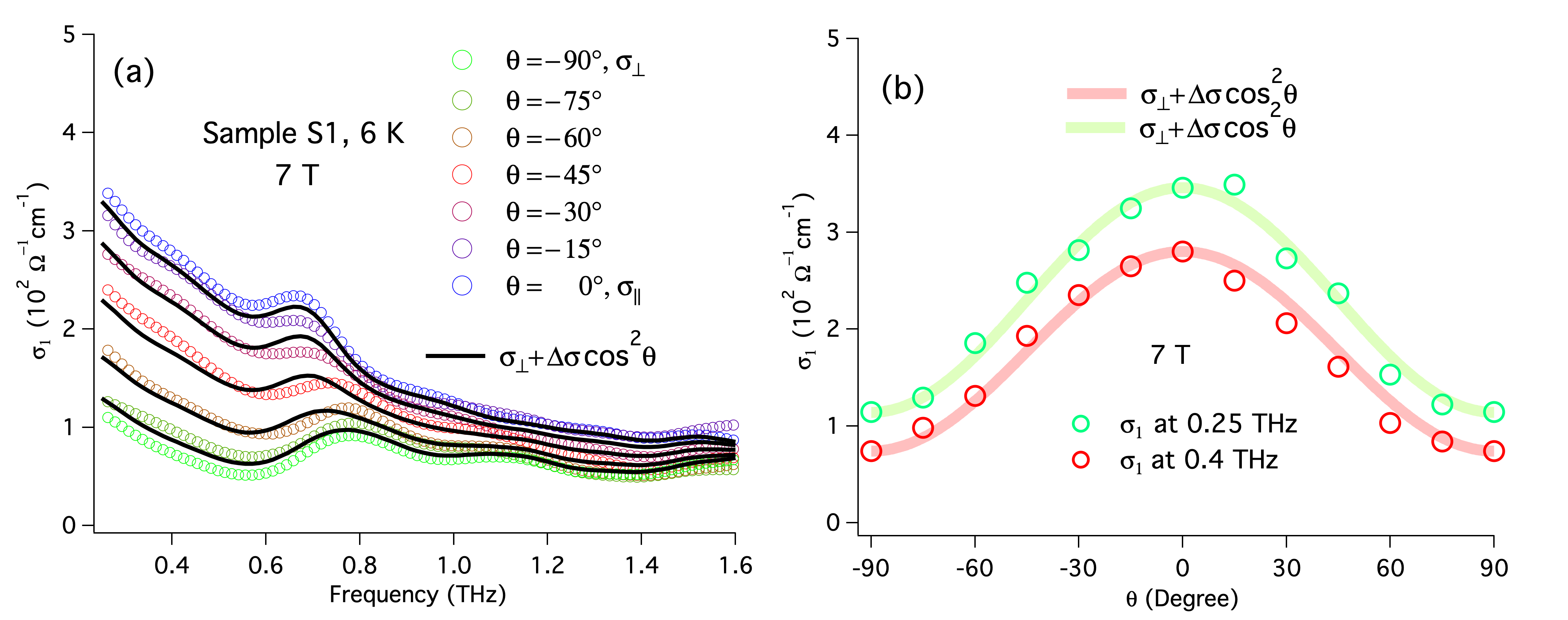}
\caption{(a) Longitudinal THz magneto-conductivity at 7 T with different terahertz polarization angles. The black curves are the fittings to conductivity with polarization angles of $-$75$^\circ$, $-$60$^\circ$, $-$45$^\circ$, $-$30$^\circ$, $-$15$^\circ$. (b) Longitudinal THz magneto-conductivity at 0.25 THz and 0.4 THz as functions of polarization angles. The green and red curves are simulations by using Eq. 1. Here $\Delta\sigma$ = $\sigma_{\parallel}-\sigma_{\perp}$.  }
\label{SI_angle_fit_s1}
\end{figure*}

\clearpage

\begin{figure*}[htp]
\includegraphics[clip,width=6.5in]{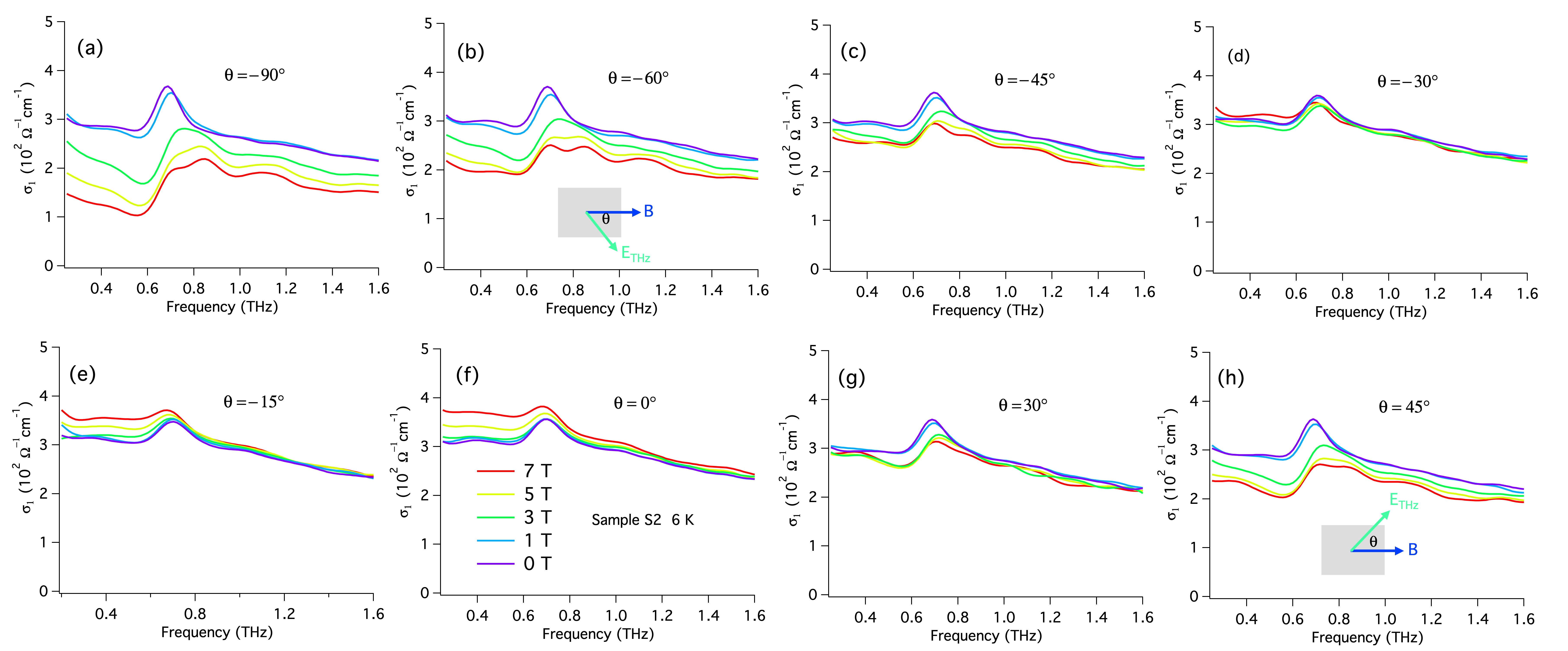}
\caption{Longitudinal THz magneto-conductivity of Sample S2 at the polarization angle of (a) $-$90$^\circ$, (b) $-$60$^\circ$, (c) $-$45$^\circ$, (d) $-$30$^\circ$, (e) $-$15$^\circ$, (f) 0$^\circ$, (g) 30$^\circ$, (h) 45$^\circ$. The data are taken at 6 K. {\bf B} is parallel to ($\overline{11}2$).}
\label{SI_angle_dep_s2}
\end{figure*}

\begin{figure*}[htp]
\includegraphics[clip,width=6in]{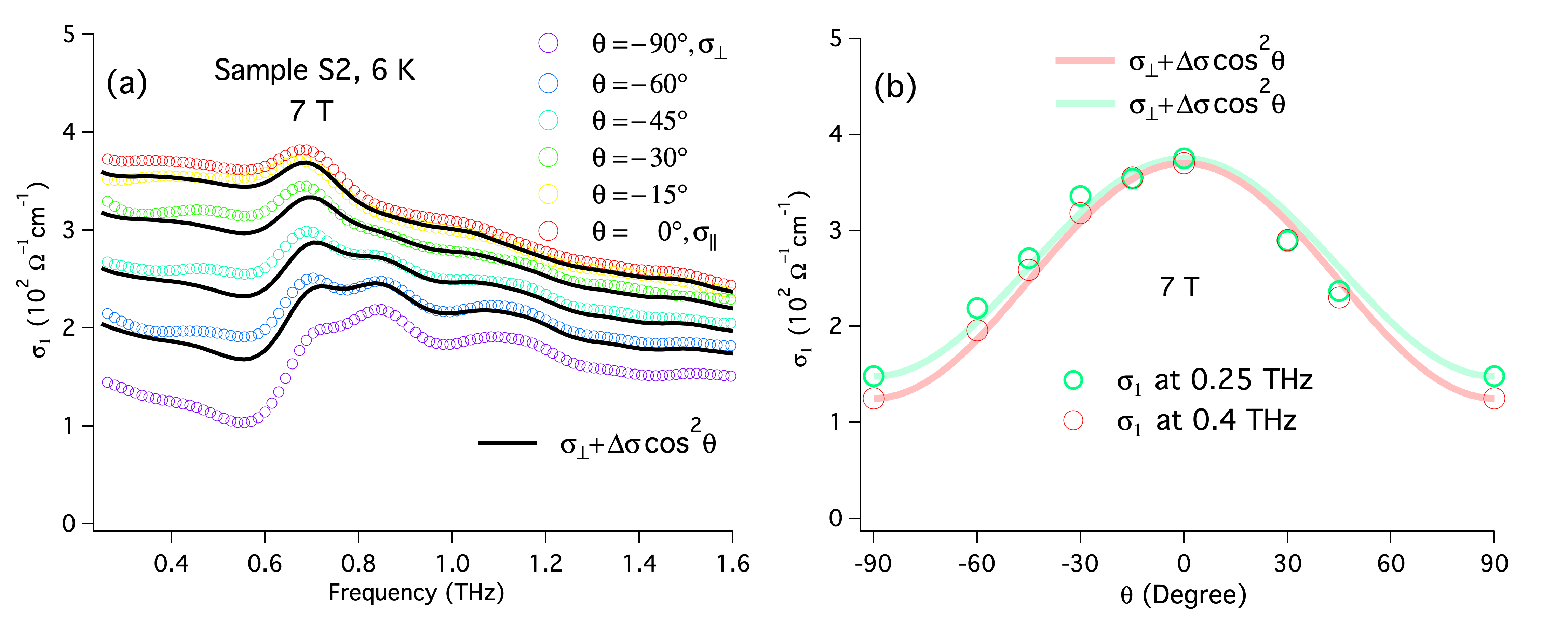}
\caption{(a) Longitudinal THz magneto-conductivity at 7 T with different terahertz polarization angles. The black curves are the fittings to conductivity with polarization angles of $-$60$^\circ$, $-$45$^\circ$, $-$30$^\circ$, $-$15$^\circ$. (b) Longitudinal THz magneto-conductivity at 0.25 THz and 0.4 THz as functions of polarization angles. The green and red curves are simulations by using Eq. 1. Here $\Delta\sigma$ = $\sigma_{\parallel}-\sigma_{\perp}$. }
\label{SI_angle_fit_s2}
\end{figure*}

\clearpage 

\begin{figure*}[htp]
\includegraphics[clip,width=7in]{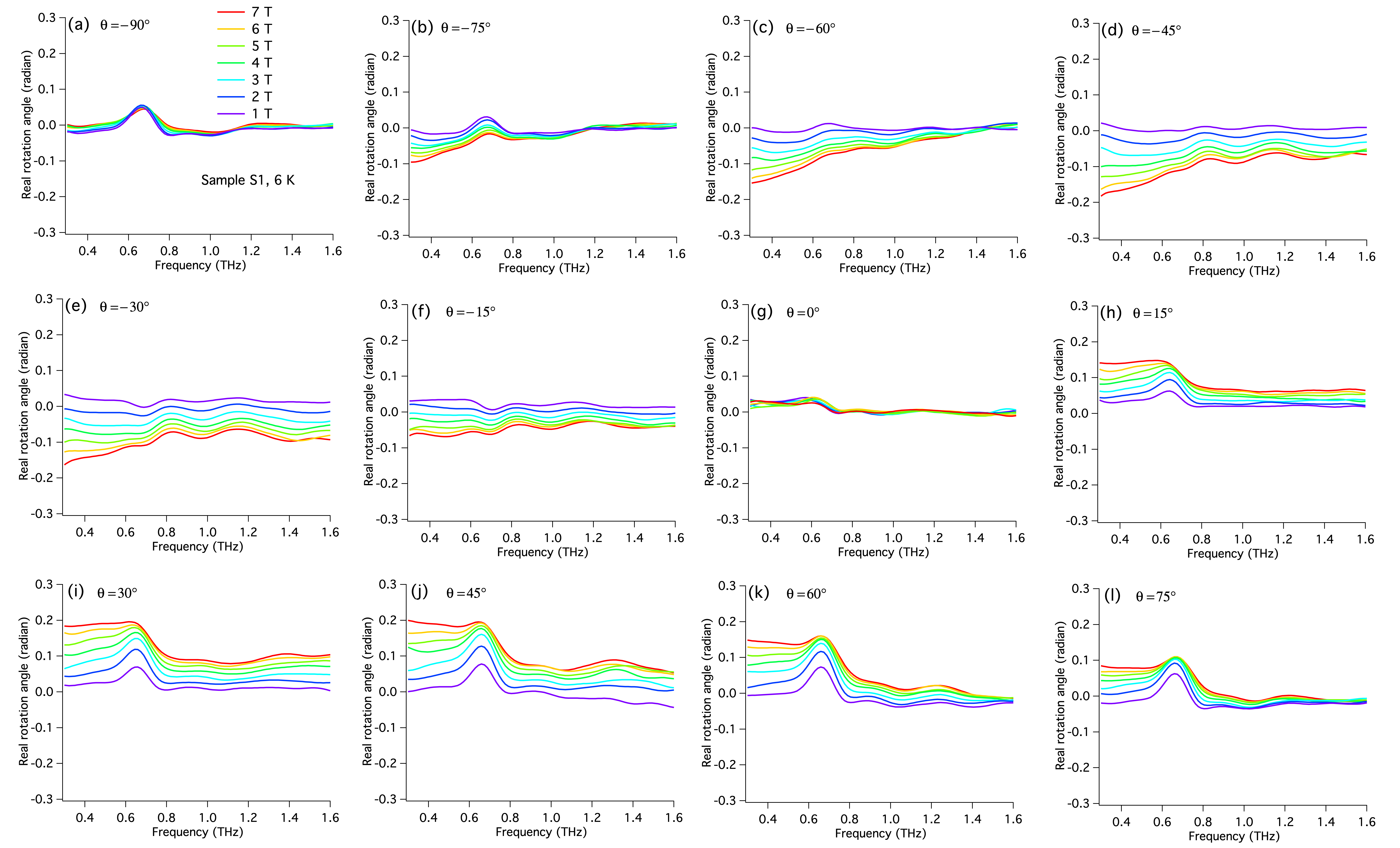}
\caption{Real rotation angle of Sample S1 at the polarization angle of (a) $-$90$^\circ$, (b) $-$75$^\circ$, (c) $-$60$^\circ$, (d) $-$45$^\circ$, (e) $-$30 $^\circ$, (f) $-$15$^\circ$, (g) 0$^\circ$, (h) 15$^\circ$, (i) 30$^\circ$, (j) 45 $^\circ$, (k) 60$^\circ$, (l) 75$^\circ$. The data are taken at 6 K. {\bf B} is parallel to ($\overline{1}$10). }
\label{SI_rotation_s1}
\end{figure*}

\begin{figure*}[htp]
\includegraphics[clip,width=3.4in]{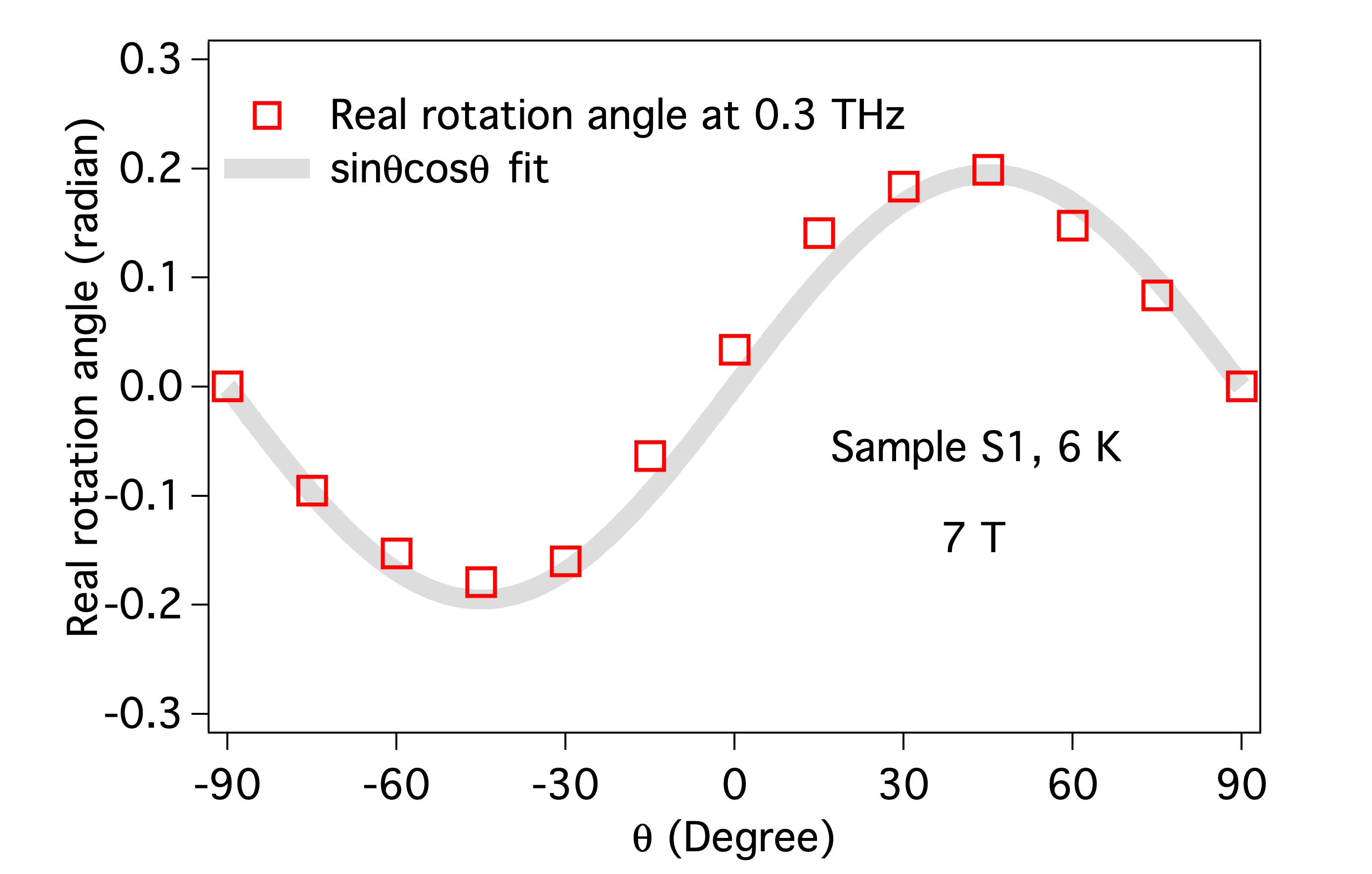}
\caption{Real rotation angle at 0.3 THz of Sample S1 as a function of THz polarization angle $\theta$. In principle, the real rotation angle is proportional to planar Hall conductivity $\sigma_{xy}$: $\varphi$ $\sim$ $\sigma_{xy}$. The gray curve is a sin$\theta$cos$\theta$ fit to the data. The data are taken at 6 K and 7 T. {\bf B} is parallel to ($\overline{1}$10). }
\label{PHE_S1}
\end{figure*}

\clearpage 

\begin{figure*}[htp]
\includegraphics[clip,width=7in]{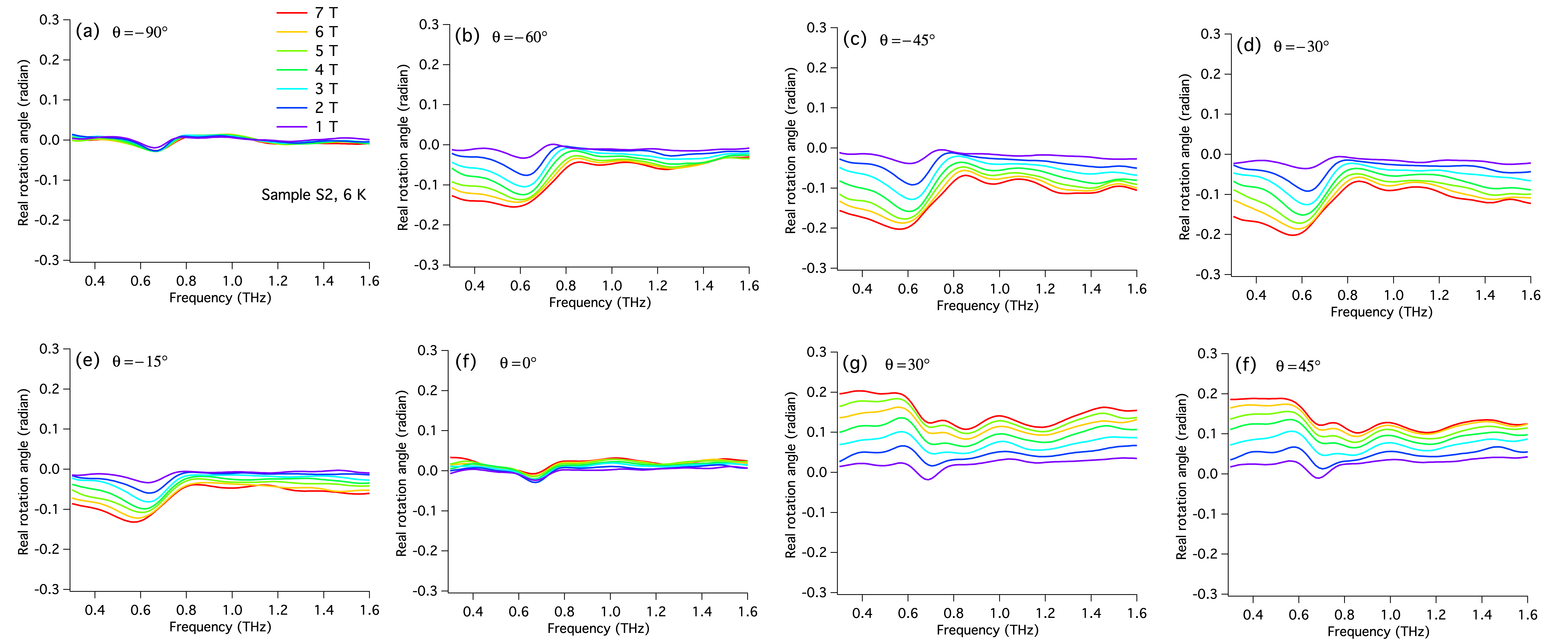}
\caption{Real rotation angle of Sample S2 at the polarization angle of (a) $-$90 $^\circ$, (b) $-$60$^\circ$, (c) $-$45$^\circ$, (d) $-$30$^\circ$, (e) $-$15$^\circ$, (f) 0$^\circ$, (g) 30$^\circ$, (h) 45$^\circ$. The data are taken at 6 K. {\bf B} is parallel to ($\overline{11}2$).}
\label{SI_rotation_s2}
\end{figure*}

\begin{figure*}[htp]
\includegraphics[clip,width=3.4in]{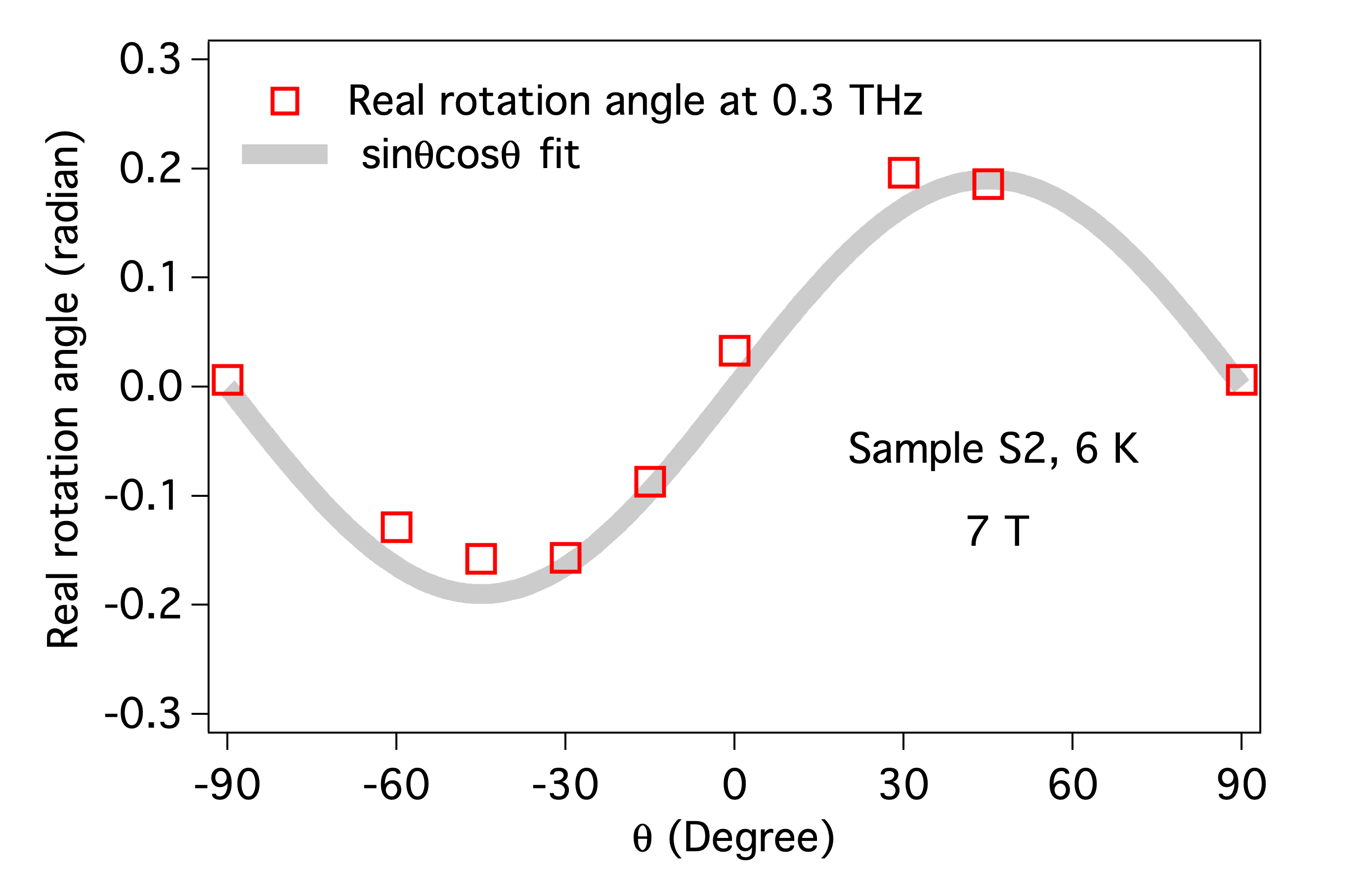}
\caption{Real rotation angle at 0.3 THz of Sample S2 as a function of THz polarization angle $\theta$. In principle, the real rotation angle is proportional to planar Hall conductivity $\sigma_{xy}$: $\varphi$ $\sim$ $\sigma_{xy}$. The gray curve is a sin$\theta$cos$\theta$ fit to the data. The data are taken at 6 K and 7 T. {\bf B} is parallel to ($\overline{11}2$). }
\label{PHE_S2}
\end{figure*}

\clearpage

\section{6. I\lowercase{n-plane anisotropy of terahertz conductivity } }


In this section, we discuss the in-plane zero field conductivity anisotropy of Cd$_3$As$_2$ films. Our films are (112) oriented and can in principle show in-plane anisotropies.  In practice these anisotropies are small in zero field.

In Fig. \ref{anisotropy} we show the terahertz conductivity of sample S1 and S2 with different polarization angles of terahertz electric field. The polarization configuration of terahertz pulse is shown by the schematic inset to Fig. \ref{anisotropy}(b). The data shown below are taken at 6 K and zero magnetic field. One can see in both sample S1 and sample S2, from $\theta$ = 0$^\circ$ to $\theta$ = 90$^\circ$, the profile of terahertz conductivity does not exhibit notable changes.  The in-plane conductivity anisotropy of both samples is within 10\%.  A weak in-plane conductivity anisotropy could be inferred from the data.

\begin{figure*}[htp]
\includegraphics[clip,width=5.4in]{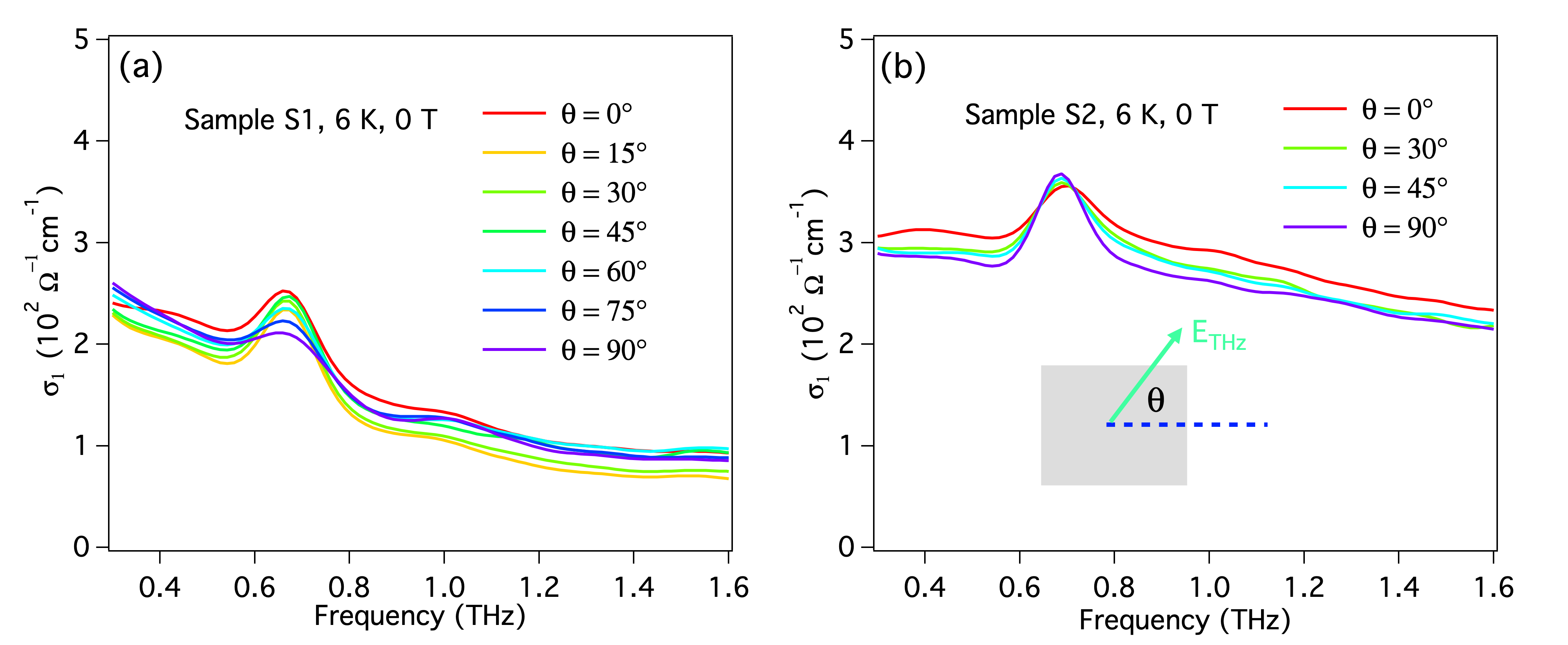}
\caption{Zero-field real part of terahertz conductivity with different terahertz electric field polarization angle of (a) sample S1, and (b) sample S2. The polarization angle $\theta$ spans from 0$^\circ$ to 90$^\circ$ and the configuration is shown by the schematic in Fig. \ref{anisotropy}(b). The data are taken at 6 K and 0 T. }
\label{anisotropy}
\end{figure*}

\clearpage

\section{7. M\lowercase{odel} H\lowercase{amiltonian and estimating the relaxation rates}  }

Similar to other semiconductors with related structure, the low energy electronic structure of Cd$_3$As$_2$ is mostly determined by near-$E_F$ Cd-5s and As-4p states.  Two features are necessary for symmetry protected Dirac nodes.  First, band inversion is essential in that the s-orbitals must be lower in energy that the p-orbitals.  Here the vacancy ordering associated with Cd is important as it is believed that a hypothetical structure without vacancies would not have the robust band inversion necessary for the Dirac semimetal state \cite{plenkiewicz1984role,Cd2As3_LDA13}.  Second, a crossing is ensured and protected along the $\Gamma - Z$ line in the Brillouin zone as the two bands belong to different representations (as distinguished by C$_4$ rotational symmetry around the k$_z$ axis) that prohibits hybridization between them in special directions in momentum space.

To understand the band structure, one can start from the standard eight-band Kane model used to model semiconductors \cite{kane1957band} and  introduce additional terms to take into account spin-orbit coupling and the tetragonal symmetry of Cd$_3$As$_2$ \cite{wang2013three,cano2017chiral}.  The Cd-5s and As-4p states with SOC can be written as states with definite angular momentum $J$ and $J_z$ e.g. $
|S_{J=\frac{1}{2}}, J_z$=$ \pm \frac{1}{2} \rangle,  |P_{J=\frac{3}{2}}, J_z $=$ \pm \frac{3}{2} \rangle,  |P_{J=\frac{3}{2}}, J_z  $=$ \pm \frac{1}{2} \rangle,$ $  |P_{J=\frac{1}{2}}, J_z  $=$ \pm \frac{1}{2} \rangle $.  In the tetragonal crystal symmetry, the total angular momentum J is no longer good quantum number and $|P_{J=\frac{3}{2}}, J_z = \pm \frac{1}{2} \rangle $ and $ |P_{J=\frac{1}{2}}, J_z = \pm \frac{1}{2} \rangle $ mix and move to high energies.   However the other two orbitals remain eigenstates, and a minimal effective low energy theory can be constructed with only them.   The  resulting 4 band Hamiltonian (with a basis $  |S_{J=\frac{1}{2}}, J_z$=$  \frac{1}{2} \rangle,  |P_{J=\frac{3}{2}}, J_z $=$  \frac{3}{2} \rangle, |S_{J=\frac{1}{2}}, J_z$=$ - \frac{1}{2} \rangle,  |P_{J=\frac{3}{2}}, J_z $=$ - \frac{3}{2} \rangle$) for an inversion symmetric tetragonal system within the $\bf{k} \cdot \bf{p}$ approximation is

\begin{eqnarray*}
  H_{\Gamma}({\pmb{k}})&\;=&\epsilon_0(\pmb{k}) +\left(\begin{array}{cccc}
      M(\pmb{k}) & Ak_{+} & 0 & B^{*}(\pmb{k}) \\
      Ak_{-} & -M(\pmb{k}) &B^{*}(\pmb{k})  & 0 \\
      0 &B(\pmb{k})  & M(\pmb{k}) & -Ak_{-}\\
      B(\pmb{k}) & 0 & -Ak_{+} & -M(\pmb{k})
\end{array}\right).
\label{Hamiltonian}
\end{eqnarray*}
The matrix elements are generic values constrained only by symmetry in the most general case, but here can be determined by experimental measures of the band structure of Cd$_3$As$_2$.   Consistent with the tetragonal symmetry one can parametrize $\epsilon_0({\pmb{k}})=C_{0}+C_{1}k_z^2+C_{2}(k_x^{2}+k_y^2)$  and $M(\pmb{k})=M_{0}-M_{1}k_z^2-M_{2}(k_{x}^{2}+k_y^{2})$.  Inverted bands correspond to $M_0, M_1, M_2 < 0$.  Here $k_{\pm}=k_{x}\pm ik_{y}$.  Diagonal terms of this matrix must be even under parity and hence even functions of momentum.  Off-diagonal terms of the $2\times 2$ blocks couple states of opposite parity and to lowest order are linear in momentum.  For an inversion symmetric tetragonal system, similarly the leading order form for $B(\pmb{k}) $ in the $k_x-k_z$ plane is $ B_3 k_z k_x^2$.  The terms that are zero on the outside of the matrix are constrained to be zero in inversion and time-reversal symmetric systems\footnote{Note that the off-diagonal terms of the Hamiltonian are constrained in different ways depending on the symmetry that protects the Dirac node.   For instance, the 3rd order terms that allow mixing of chiralities can be different for the C$_3$ symmetry that protects the nodes in Na$_3$Bi and the C$_4$ symmetry that protects the nodes in Cd$_3$As$_2$.  This is true for both the terms that couple diagonal entries within a $2\times 2$ block and the terms that couple $2\times 2$ blocks. See Ref. \cite{cano2017chiral,Na3Bi_LDA_2012} for examples. However these different symmetries give the same result for the $k_x-k_z$ plane mostly considered here.}.  We discuss the effects of adding a magnetic field below.  In this form, it is only the off-diagonal $B$ term that mixes chiralities and among other aspects has the important effect of allowing rotationally invariant impurities to scatter between nodes.   Below, we will use the fact that since the $B(\pmb{k}) $ terms are small near the nodes the $4\times4$ matrix can be almost decoupled into two $2\times2$ blocks each describing isolated Weyl points with distinct isospin.  However, with the leading order term for $B(\pmb{k}) $ the energy dispersion in the $k_x-k_z$ plane is

\begin{equation}
E(\pmb{k}) = \epsilon_0(\pmb{k})  \pm   \sqrt{    M(\pmb{k})^2    + A^2 k_x^2 +  B_3^2   k_z^2 k_x^4   }     
\end{equation}  
which gives a pair of fourfold degenerate Dirac points at $\pmb{K}_D = (0,0,\sqrt{ \frac{M_0}{M_1}   } )$.

Using the above formalism one can make estimates for both the large momentum scattering between nodes $1/\tau_v$ as well as the intravalley inter-isospin scattering at the same node $1/\tau_i$.  Following Ref. \cite{Parameswaran14a}, one can assume a random distribution of impurities, which can be modeled by a smooth potential.   An assumption (possibly a big one, see below) is that the potential is spherically symmetric and has a scale that is large compared to the lattice constant, such that crystalline point-group symmetries are not broken locally.   In reality, impurities placed away from inversion centers are likely to allow mixing between isospins in a fashion different than considered here.  If one assumes a screened Coulomb potential $v(\vec{q}) = \frac{v_0}{q^2 + k_{sc}^2}$ with a screening length ($1/k_{sc}$) of order the Fermi wavelength, using Fermi's golden rule one may estimate for the relaxation time \cite{Parameswaran14a} for intra-node (e.g. small momentum $|\vec{q}| \ll |k_F|$) intra-isospin scattering

\begin{equation}
\frac{1}{\tau_n} \approx \frac{ g(\epsilon_F) n_{imp} }{\pi \hbar}    |v(0)|^2.
\end{equation}
Here $ g(\epsilon_F) $ is the Fermi energy density of states and $n_{imp}$ is the impurity concentration.

For large momentum, inter-valley scattering, where one has $k_F \ll |K_D|$ one can calculate \cite{Parameswaran14a} in a similar fashion to get 

\begin{equation}
\frac{1}{\tau_v}  \approx  \frac{ g(\epsilon_F) n_{imp} }{\pi \hbar}    |v(2K_D)|^2 =   \frac{1}{\tau_n}  \frac{    |v(2K_D)|^2}{ |v(0)|^2   }    \sim  \frac{1}{\tau_n}  \Big(  \frac{k_F}{2K_D }  \Big)^4.
\end{equation}
Here the Dirac node momentum $K_D$ position is measured with respect to the zone center and $k_F$ is measured with respect to the node position.

The calculation of the intravalley inter-isospin scattering is slightly more involved.   Following Ref. \cite{Parameswaran14a} one can first diagonalize the Hamiltonian in the limit where  $B(k) = 0 $  and valleys are chirality eigenstates for all $\pmb{k}$.   First-order perturbation theory in  $B(k) $ allows chiralitites to be mixed due quadratic curvature corrections.    Note that within this picture chiralities are still eigenstates as $k \rightarrow K_D$, but are progressively mixed at momenta away from the node.   Repeating the Fermi's golden rule calculation using the perturbed eigenstates and ignoring any angular dependence to the scattering or shape of the Fermi surface or velocity, one finds

\begin{equation}
\frac{1}{\tau_i}  \approx  \frac{ g(\epsilon_F) n_{imp} }{\pi \hbar}    |v(0)|^2  \Big ( \frac{ B_3 K_D k_F}{ 2 \hbar v_F} \Big )^2    =  \frac{1}{\tau_n}   \Big ( \frac{ B_3 K_D k_F}{ 2 \hbar v_F} \Big )^2.
\end{equation}
 
Fermi surface and dispersion anisotropies will give only order one corrections to this result.  One can see from this analysis that it is part of the quadratic correction to the dispersion in the direction perpendicular to the high symmetry  $\Gamma - Z$ line (see further discussion below) that leads to mixing between chiralities.   The quadratic dispersion along the  $\Gamma - Z$ line has no effect in this regard.  The above expression can be put in alternative form based on known observables.  In a massless Dirac system the effective cyclotron mass $m^*$ is $\hbar k_F/v_F$ and $E_F = \hbar v_F k_F$.   Therefore one can also write

\begin{equation}
\frac{1}{\tau_i}  \approx      \frac{1}{\tau_n}   \Big ( \frac{ B_3 K_D k_F^2}{ 2 E_F} \Big )^2 = \frac{1}{\tau_n}   \Big ( \frac{ B_3 K_D m^*}{ 2\hbar^2} \Big )^2 .
\end{equation}

\section{8. E\lowercase{stimating the scattering rates from the measured band structure}}

Within the context of this simple theory, one can attempt to estimate the essential parameters $K_D$, $B_3$, $v_F$, $m^*$, and $k_F$ from the measured band structure from our own data (for $k_F$) and  angle-resolved photoemission (for the rest) to estimate values of $1/\tau_v$ and $1/\tau_i$ in the context of the above theory.  Unfortunately photoemission datasets of high enough quality do not yet exist to determine these parameters uniquely.  This is primarily due to challenges associated with the 3D nature of the material, and the fact that the natural cleavage plane of this compound is (112) and there is relatively poor $k_z$ resolution.

The separation of Dirac points is most easily measured, but even here there are challenges.  From Ref. \cite{borisenko2014experimental}, one can estimate that $K_D$ is found approximately 1/3 of the way towards $Z$ along the $\Gamma-Z$ line.   This is $0.04  \pm 0.02 \AA^{-1}$.  This value is consistent with bandstructure calculations \cite{wang2013three,borisenko2014experimental,neupane2014observation}. 

\begin{figure}[ht] 
\includegraphics[clip,width=3.5in]{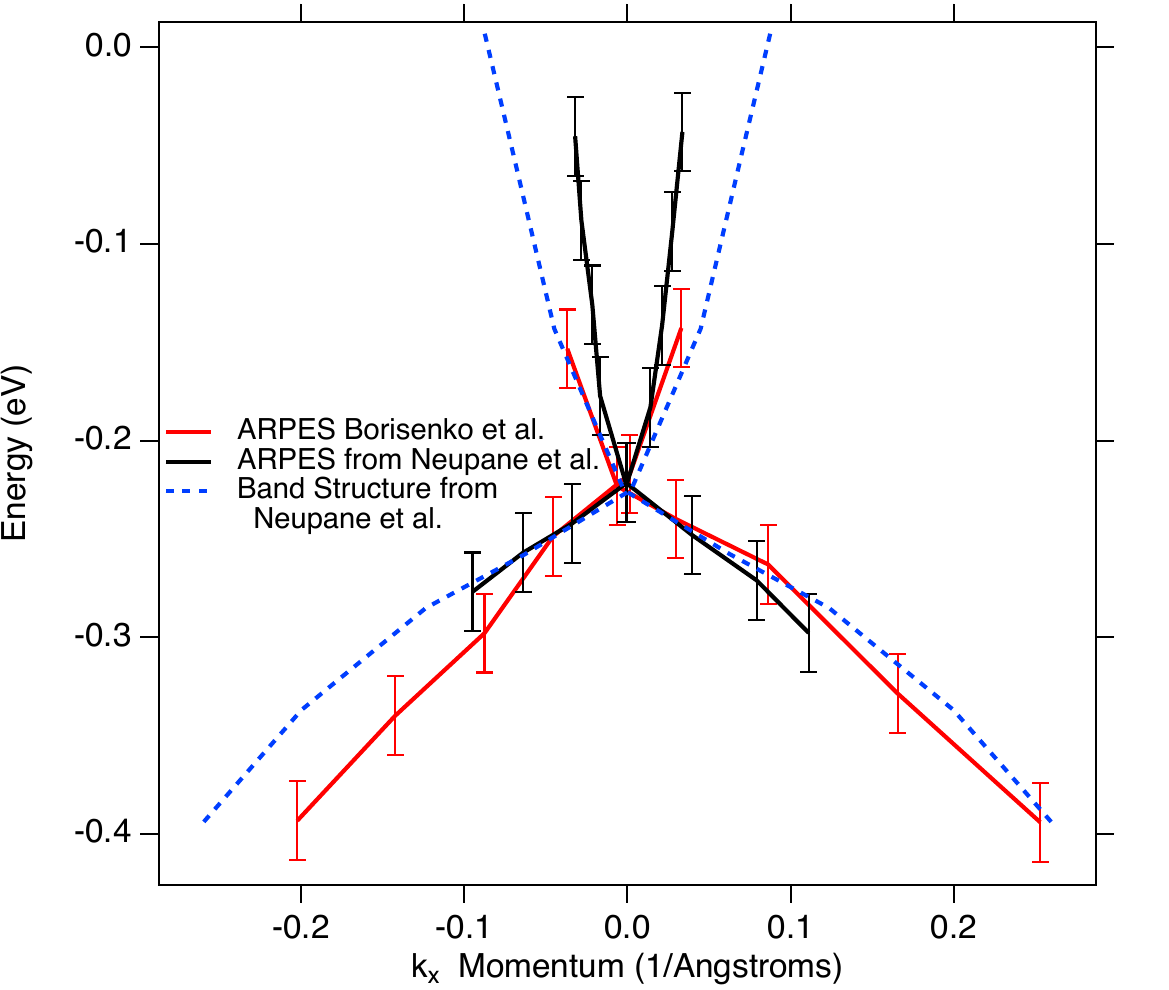}
\caption{Estimation of band dispersions in Cd$_3$As$_2$ extracted from ARPES \cite{borisenko2014experimental,neupane2014observation}.  Also shown is extracted dispersions from the band structure calculations in Ref. \cite{neupane2014observation}. Data presented along the $k_x$ direction and taken at a $k_z$ position that cuts through the Dirac node.  Here $E_F$ of the data from Ref. \cite{borisenko2014experimental} has been shifted by 85 meV downwards to match the data of Ref. \cite{neupane2014observation}.}
\label{Dispersions}
\end{figure}

The quadratic correction $B(k) = B_3   k_z k_x^2$ that mixes chiral sectors is more difficult to estimate.   As discussed above, it manifests in quadratic corrections to the dispersion perpendicular to the $\Gamma - Z$ line, but as can be see from Eq. 1, a quadratic contribution to the dispersion perpendicular to $\Gamma - Z$ can enter in $\epsilon_0$, $M(\pmb{k})$, or $B(\pmb{k})$.   The contribution of $\epsilon_0$ can be separated from the other two because it gives the same sign energy contribution to both positive and lower energy branches, whereas the contribution from $M(\pmb{k})$ or $B(\pmb{k})$ give opposite energy contributions to upper and lower branches.   In principle, $M(\pmb{k})$ and $B(\pmb{k})$ could  be isolated from each other by doing a systematic study where the $k_x$ dispersion is probed at different $k_z$.  $M(\pmb{k})$ would give a contribution where its quadratic coefficient (in $k_x$) would be independent of $k_z$ whereas $B(\pmb{k})$'s quadratic coefficient would be proportional to $k_z$.   Unfortunately, ARPES data with high enough quality to perform this analysis does not exist.

Therefore to make further analysis we assume that the $k_x$ coefficient associated with the $M(\pmb{k})$ term (e.g. $M_2$) is zero.  This allows us to put an upper bound on $B_3 K_D$, which is sufficient for our purposes.  We fit the experimental dispersion curves extracted from ARPES (Fig. \ref{Dispersions}) to a simplified version of Eq. 1 \cite{borisenko2014experimental,neupane2014observation}.   Also shown (Fig. \ref{Dispersions}) are band structure calculations also extracted from Ref. \cite{neupane2014observation}.   We fit to the reduced expression 

\begin{equation}
    E(\pmb{k}) = C_{0}+C_{1}K_D^2+C_{2}k_x^{2}  \pm   \sqrt{      A^2 k_x^2 +  B_3^2   K_D^2 k_x^4   }.
\end{equation}

Here there is some variability in fit parameters from different data sets, but the shape of the dispersion gives signatures of both finite $C_2$ and $B_3 K_D$.  With only $C_2$ the dispersions would be simple parabolas crossing at the Weyl point exhibiting an asymmetry in the upper and lower branches.   With only $B_3 K_D$ the upper and lower branches would be symmetric around the touching point.  The ARPES shows that both contributions are present.  From these fits we determine $A = 1.2 \pm 0.2$ eV $\cdot \AA$ and $B_3 K_D = 30 \pm 15$ eV $\cdot \AA^2$.   It is important to note the parameter $A$ does not give the velocity at $E_F$ directly as there are substantial quadratic corrections.

We can estimate $k_F$ from a measure of the zero-field Drude spectral weight and the cyclotron resonance.   This is discussed below.  A cyclotron mass of $m^* = 0.03 m_e$ was determined in our previous work \cite{cheng2019large}.  For sample S1, we find then $k_F$ of 0.0135 $\AA^{-1}$, which gives the valley scattering rate $1/\tau_v = 0.0008$ $ 1/\tau_n$.  This is much smaller than observed.  Using the above value for $B_3 K_D$ and the measured value for the cyclotron mass, one estimates an upper bound on the value of the isospin scattering rate $1/\tau_i = 0.0035$ $ 1/\tau_n$, which is about four times larger than $1/\tau_v$, but also much smaller than observed.

We caution that the scheme used here relies on first-order perturbation theory for analysis, but the disagreement is so large that we believe the whole scheme may need to be revised. For instance it is possible that scatterings are involved that couple chiralities at $O(k)$ instead of $O(k^2)$ due to impurities that break local inversion.  An approximation we used was that impurities had scales large as compared to the lattice constant. In this regard one should keep in mind that the unit cell in Cd$_3$As$_2$ is extremely large and associated with cadmium vacancy ordering. It is likely that the main source of disorder is defects associated with this ordering. In which case the potential will not be smooth on the scale of the unit cell.

\section{9. T\lowercase{he effects of broken rotational symmetry by $x-y$ oriented field}}

As discussed above and in the main text, the degeneracies of Dirac semimetals are protected by point group symmetries.   Therefore it is important to consider the role of magnetic field applied in directions other than the $z$ direction which preserves the $C_{4z}$ symmetry of Cd$_3$As$_2$.  The modified band structure in the presence of magnetic field can be evaluated by including only the Zeeman term.  It takes the form

\begin{equation}
    H_Z = - \mu_B (g_s \mathbf{J}_s + g_p \mathbf{J}_p) \cdot \mathbf{B}.
    \label{Zeeman}
\end{equation}

One can note that since $(J_{s,p})_{x,y}$ can be written in the form of raising and lower operators, they only couple like orbitals with $ \Delta m_j = \pm 1$ and hence $(J_p)_{x,y}$ only couples our basis states to states outside the low energy manifold (e.g. it does not mix the $|P_{J=\frac{3}{2}}, J_z $=$ \pm \frac{3}{2} \rangle$ states that are in the low energy manifold).  Therefore the principal effect of an $x$ or $y$ field is on the $|S_{J=\frac{1}{2}}, J_z $=$ \pm \frac{1}{2} \rangle$ states.  Ref. \cite{cano2017chiral} included the effects of a $x$ or $y$ field on the $|P_{J=\frac{3}{2}}, J_z $=$ \pm \frac{3}{2} \rangle$ states perturbatively on their energies, but for our purposes the essential physics can be found by including the effect of only the four matrices in Eq. \ref{Zeeman}

\begin{eqnarray*}
 (J_s)_z = \frac{1}{2} \left(\begin{array}{cccc}
      1 & 0 & 0 & 0 \\
      0& 0 &0  & 0 \\
      0 & 0  & -1  & 0 \\
    0 & 0 &0   & 0
\end{array}\right), \;
(J_s)_x = \frac{1}{2} \left(\begin{array}{cccc}
      0 & 0 & 1 & 0 \\
      0& 0 &0  & 0 \\
      1 & 0  & 0  & 0 \\
    0 & 0 &0   & 0
\end{array}\right), \;
(J_s)_y = \frac{1}{2} \left(\begin{array}{cccc}
      0 & 0 & i & 0 \\
      0& 0 &0  & 0 \\
      -i & 0  & 0  & 0 \\
    0 & 0 &0   & 0
\end{array}\right), \;
(J_p)_z = \frac{3}{2} \left(\begin{array}{cccc}
      0 & 0 &  & 0 \\
      0& 1 &0  & 0 \\
      0 & 0  & 0  & 0 \\
    0 & 0 &0   & -1
\end{array}\right).
\label{Jmatrix}
\end{eqnarray*}

When the magnetic field is along the $z$ direction all band crossings are protected by $C_{4z}$ symmetry.  It is clear from the form of the above couplings that a $z$ axis oriented field does not cause any additional gappings.   Ref. \cite{cano2017chiral} showed this explicitly for Cd$_3$As$_2$.  Depending on the value of the magnetic field there are between four and eight Weyl points.   Crossings between the $J_z = \pm   \frac{3}{2}$ and $J_z = \pm   \frac{1}{2}$ bands are Weyl points.  Crossings between the $J_z = \pm   \frac{3}{2}$ and $J_z = \mp   \frac{1}{2}$ bands are double Weyl points\footnote{As shown in Ref. \cite{cano2017chiral} double Weyl points exhibit a modified chiral anomaly with the form of all transport coefficients staying essentially the same.  The only changes is that the response coefficients are proportional to the square of the Chern number and the form of the density of states changes}.  The latter are not robust to small changes in magnetic field and will spit into single Weyl nodes upon misaligning the field from $z$.  However, it is important to note that since the Weyl nodes are topological objects they do not annihilate by such field misalignments despite the fact that such a field breaks the rotational symmetry.

Regarding the effects of magnetic field, it is also important to note that even putting the above considerations aside, the energetic changes from the laboratory fields are small even with the estimates of a large $g_s$ (of order 16 for the $|S_{J=\frac{1}{2}}, J_z $=$ \pm \frac{1}{2} \rangle$ states \cite{narayanan2015linear}).   At our largest laboratory field (7 T) the $\mathbf{J}_s$ terms in Eq. \ref{Zeeman} gives a contribution of order 3.25 meV, which is very modest on the scale of our Fermi energies ($\sim 46$ meV).   The contribution from the $\mathbf{J}_p$ terms are much smaller.

The effects of breaking the rotational symmetry by an $x$ directed magnetic field on the scattering between isospin species can be estimated in a similar fashion as we did above when considering the effect of the quadratic curvature corrections.  It happens in much the same fashion as that case by allowing a small amount of mixing between chiral channels.  We first diagonalize the Hamiltonian in the limit where  $B(k) = 0 $ and $H_Z = 0$ and valleys are chirality eigenstates for all $\pmb{k}$.   Then we do first-order perturbation theory, but this time in the applied field.   Different from the case of quadratic corrections which vanished near the node, the Zeeman field is finite for all energies and hence the perturbation theory is only well defined when it can be taken to be a small as compared to the Fermi energy.   Therefore unlike the case of quadratic correction, this analysis and a chiral contribution to the transport will only occur for x-axis fields at finite $E_F$.  Again neglecting any anistropies in the matrix elements (which only change the result by numbers of order unity), one finds via Fermi's golden rule that the additional scattering incurred due to chiral mixing is 

\begin{equation}
\frac{1}{\tau_{B_x}}  \approx   \frac{1}{\tau_n}   \Big ( \frac{ g_s \mu_B B }{ 4 E_F} \Big )^2.
\end{equation}

This is estimated to be about 0.00023$ \frac{1}{\tau_n}$ for our largest 7 T field.   This is four times less than even the estimates for the valley chiral scattering, which is again much smaller than the values observed.  Moreover, this effect -- if observed -- would go like the square of the magnetic field.   This is also inconsistent with our data.   Therefore we do not believe chirality mixing due to an x-axis magnetic field is a large contribution to the chiral relaxation rates.

\section{10. E\lowercase{stimating the} F\lowercase{ermi energy}}

In the sample S1 the zero-field Drude fit, gives a Drude plasma frequency $\omega_p$/2$\pi$ of 21.2 THz. The charge density $n$ and Drude plasma frequency $\omega_p$/2$\pi$ are related by the formula: $\omega_p^2$ = $n$e$^2$/m$^*\epsilon_0$. From our previous study of cyclotron energy, the effective mass m$^*$ of Dirac fermions is estimated to be 0.03 $m_e$ in sample S1, where $m_e$ is the free electron mass \cite{cheng2019large}. When the Fermi energy is not large, it is reasonable to assume that the linear dispersion of 3D Dirac cones is isotropic. Then we could directly relate the charge density $n$ and Fermi velocity k$_F$ by Luttinger's theorem: $n$ = 2k$_F^3$/3$\pi^2$. Here the valley and isospin degeneracy of the Dirac cones have been accounted for. In this fashion, the Fermi vector k$_F$ in sample S1 could be estimated to be 0.013 \AA$^{-1}$ and the Fermi energy E$_F$ = $\hbar^2$k$_F^2$/m$^*$. After inputting all parameters, E$_F$ is estimated to be 46 meV. For the sample S2 that showed a weaker magnetic field effect, one can perform the same procedures and its k$_F$ and E$_F$ are extracted to be 0.018 \AA$^{-1}$ and 118 meV respectively.


\end{widetext}

\end{document}